\documentclass[final,authoryear,2p,12pt]{elsarticle}
\usepackage{geometry}                
\usepackage{graphicx}
\usepackage{epstopdf}
\usepackage{booktabs}
\usepackage{natbib}
\usepackage{fancyhdr}

\usepackage{hyperref}
\usepackage{amsthm}
\usepackage{cleveref}

\usepackage{amsmath}
\usepackage{amsfonts}
\usepackage{multirow}
\usepackage{lscape}
\usepackage{bbm}
\usepackage{bm}
\usepackage{color}

\usepackage{ifthen}
\usepackage[latin1]{inputenc}
\usepackage{titletoc}
\usepackage{eso-pic}
\usepackage{float}

\usepackage{amssymb}
\usepackage{xcolor}

\newtheorem{prop}{Proposition}

\newcommand{\ra}[1]{\renewcommand{\arraystretch}{#1}}

\usepackage{color,soul}
\definecolor{lightblue}{rgb}{.80,.95,1}
\sethlcolor{lightblue}
\setulcolor{red}

\begin{document}
\begin{frontmatter}

\title{Do co-jumps impact correlations in currency markets?}

\author[ies]{Jozef Barunik\corref{cor2}}
\author[ies,utia]{Lukas Vacha}

\address[ies]{Institute of Economic Studies, Charles University in Prague, Opletalova 26, 110 00 Prague, Czech Republic}
\address[utia]{Institute of Information Theory and Automation, The Czech Academy of Sciences, Pod Vodarenskou Vezi 4, 182 00 Prague, Czech Republic}
\cortext[cor2]{Corresponding author, Tel. +420(602)161710, Email address: vachal@utia.cas.cz}
\tnotetext[label1]{We are grateful to the editor Tarun Chordia and an anonymous referee for useful comments and suggestions, which have greatly improved the paper. Furthermore, we are indebted to Torben Andersen, Ionut Florescu, Giampiero Gallo, Karel Najzar, Roberto Ren\`{o}, David Veredas, and seminar participants at the Vienna-Copenhagen Conference on Financial Econometrics (Vienna, 2017), 69th European meeting of the Econometric Society (Geneva, 2016), Conference on Financial Econometrics and Empirical Asset Pricing (Lancaster, 2016), Modeling High Frequency Data in Finance (New York, 2015), Non- and Semiparametric Volatility and Correlation Model (Paderborn, 2014), and Computational and Financial Econometrics in London (London 2011) for many useful comments, suggestions and discussions. Support from the Czech Science Foundation under the GA16-14151S project is gratefully acknowledged.}

\begin{abstract}
We quantify how co-jumps impact correlations in currency markets. To disentangle the continuous part of quadratic covariation from co-jumps, and study the influence of co-jumps on correlations, we propose a new wavelet-based estimator. The proposed estimation framework is able to localize the co-jumps very precisely through wavelet coefficients and identify statistically significant co-jumps. Empirical findings reveal the different behaviors of co-jumps during Asian, European and U.S. trading sessions. Importantly, we document that co-jumps significantly influence correlation in currency markets.
\end{abstract}
\begin{keyword}
co-jumps \sep currency markets \sep realized covariance \sep wavelets \sep bootstrap
\end{keyword}
\end{frontmatter}
\textit{JEL: C14, C53, G17}

\newpage
\section{Introduction}

One of the fundamental problems faced by a researcher trying to understand financial markets is how to quantify the dependence between assets. Although commonly used correlation based measures are essential tools used to uncover the dependence structures, exogenous events resulting in idiosyncratic and systemic jumps, or co-jumps, may impact the measurements. Being equally important part of the information, co-jumps and its role need to be understood fully before making any conclusions about dependence. In this paper, we focus on estimating the effects of these exogenous events in order to see how co-jumps impact correlations in currency markets. Since correlation is covariance normalized by variance, we propose a wavelet based framework to accurately estimate total covariance, as well as disentangle the continuous from discontinuous (co-jump) part of covariation. Having the decomposition in hand, we define the continuous correlation as a measure that is not dependent on important market announcements (co-jumps) or extreme univariate shocks of the single asset (jumps). Comparing the total and continuous correlations, we answer the question how co-jumps impact correlations on currency markets. In addition, we document the co-jump, covariance, and correlation dynamics for the three main trading sessions--Asian, European and the U.S.--to determine where the dependence is being created.

Distinguishing between continuous and co-jump covariation is important for asset pricing as both parts carry different sources of risk leading to different optimal hedging strategies in asset pricing models \citep{sahaliajacod2009}. While the continuous covariation part of asset components can be well diversified in the portfolio, the presence of co-jumps implies that the construction of a hedging portfolio has to consider new constraints \citep{mancini2012identifying}. Moreover, separating the contribution of continuous and (co-)jump covariation in asset prices is crucial for investors. For example, the correlation between an asset and the stock market is an essential part of the capital asset pricing model (CAPM). Hence, an increase in total correlation due to the presence of co-jumps will increase market price risk, or beta, and an investor needs to be aware of this to be able to price this part of financial risk. 

Modeling the covariance structures has received considerable attention in the literature. With increased availability of high-frequency intraday data, the literature has shifted from parametric conditional covariance estimation toward model-free measurement. This paradigm shift from treating covariances as latent towards directly modeling ex-post covariance measures constructed from intraday data \citep{abdl2003,barndorff2004} has spurred additional interest. Although the theory is appealing and intuitive, it assumes that the observed high-frequency data represent the underlying process. Nevertheless, the real-world data contains microstructure noise and jumps, which makes drawing statistical inferences rather difficult. 

To address the presence of microstructure noise, researchers often collect sparsely sampled observations. This approach reduces the bias due to noise, but discards a very large amount of data directly. Although it is statistically implausible, the reason is based on an empirical observation of increasing biases with increasing data-collection frequency. The desire to use all available data at higher frequencies has led to a number of proposed approaches to restore consistency through subsampling, for example, \cite{zhang2005}'s two-scale realized volatility estimator. \cite{Zhang2011} generalizes these ideas to a multivariate setting and defines a two-scale covariance estimator. \cite{BN2011} achieve positive semi-definiteness of the variance-covariance matrix using multivariate kernel-based estimation. Furthermore, \cite{griffin} and \cite{sahalia2011} address microstructure noise and non-synchronous trading and propose a consistent and efficient estimator of realized covariance. \cite{ait2012analyzing} analyze the effects of microstructure noise and jumps, and \cite{varneskov2015flat} estimate quadratic covariation using a general multivariate additive noise model.

In addition to the microstructure noise, ignoring jumps and co-jumps can substantially influence the results of estimation, especially with regard to forecasting, option pricing, portfolio risk management and credit risk management \citep{jawadi2015testing}. Building on univariate jump detection\footnote{The univariate jump detection is addressed, for example,  in \cite{barndorff2006},  \cite{abd2007}, \cite{lee2008}, \cite{sahaliajacod2009}, \cite{jacod2009testing}, and \cite{novotny2015trading}, among others.}, the literature has lately focused on detecting co-jumps and multi-jumps. \cite{bollerslev2008risk} detect co-jumps in a large panel of intraday stock returns in an equally weighted portfolio. They propose a mean cross-product statistic that directly measures how closely the stocks co-move. \cite{Lahaye2011} use \cite{lee2008}'s univariate jump test to identify co-jumps, defined as jumps occurring simultaneously on different markets. They call this approach ``univariate co-jumps'' because their detection relies on univariate jump detection. In addition, \cite{mancini2012identifying} observe co-jumps via thresholding techniques. Recently, spectral techniques for co-jump detection have been employed by \cite{bibinger2015econometrics}. \cite{gilder2014cojumps} use the approach of \cite{bollerslev2008risk} to identify co-jumps at daily frequency. Because this method is not robust against disjoint co-jumps, these authors further utilize tests for intraday jumps, as described by \cite{andersen2010continuous}. \cite{boudt2015jump} propose a jump robust version of \cite{Zhang2011}'s two-scale covariance estimator. A test statistic that can explicitly identify co-jumps is proposed in \cite{gnabo2014system} and accounts for the assets' covariation, considering a co-jump as large cross product of returns with respect to local covariation. A common problem associated with this method is that it can lead to false co-jump detection when a substantially large jump occurs in only one asset. Extension to a multivariate space is proposed by \cite{caporin2014multi}, who use a formal test to detect multi-jumps in larger portfolios. Their procedure is based on comparing two types of smoothed power variations.

In this study, we contribute to the growing literature by introducing an approach based on a wavelet decomposition of stochastic processes. The main reason why we focus on wavelet analysis is its remarkable ability to detect jumps and sharp cusps even if covered by noise \citep{donoho,wang95}. Several authors have used these results to improve the jump estimation \citep{fanwang2008,barunik,barunikjwg,xue2014jump}. The reported improvements originate from the fact that wavelets are able to decompose noisy time series into separate time-scale components. This decomposition then helps to distinguish jumps from continuous price changes, and microstructure noise effects as wavelet coefficients decay at a different rate for continuous and jump processes. Wavelet coefficients at jump locations are larger in comparison to other observations in the. While changes in continuous price processes over a given small time interval are close to zero, changes in jumps are not. Wavelet coefficients are able to precisely distinguish between these situations, and hence locate jumps very precisely. Specifically, the first scale wavelet coefficients represents only the highest frequency, thus they can detect sharp discontinuities in the process without being influenced by other frequency components.

We step forward, and explore these features on more than one asset by providing a technique that allows for precise separation jumps and co-jumps while minimizing false co-jumps resulting from large idiosyncratic jumps. Moreover, we improve the finite sample properties of the jump and co-jump tests based on realized measures  by extending the bootstrap tests developed in a univariate setting \citep{dovonon2014bootstrapping}. The estimator we propose is based on the two-scale covariance estimator framework of \cite{Zhang2011}, and thus, it is able to utilize all available data using an unbiased estimator in the presence of noise. We test the small sample performance of the estimator in a large numerical study and compare it to other popular integrated covariance estimators under different simulation settings with varying noise and co-jump levels. The results show that our wavelet-based estimator can estimate the realized covariance from data containing microstructure noise, jumps, and co-jumps with high precision.

While we are the first to explore the usefulness of wavelet decomposition in estimating covariance and co-jumps, we view our main contribution in documenting how precisely localized co-jumps impact correlation structures in the currency market. Empirical findings reveal the behavior of co-jumps during Asian, European and U.S. trading sessions. We document how co-jumps are becoming more important part of the total correlations in currency markets as the proportion of co-jumps relative to the covariance increased during the years 2012 -- 2015. Hence, appropriately estimating co-jumps is becoming a crucial step in understanding dependence in currency markets. 

\subsection{Recent surge of frequency-based methods in financial economics}

Relying heavily on frequency domain methods, it is useful to motivate its use and position our contribution to the recent literature prior to introducing the framework we work with. Recently, there has been an important surge of studies using frequency and time-frequency based methods in finance and economics. The central idea is to decompose aggregate information in the data using filtering techniques (Fourier transform, wavelets, etc.) to capture cyclical properties. The main reason for this is that financial and economic data have cycles that remain hidden when the classical time series approach is used because it averages information on all the frequencies. Lately, several studies show that (frequency) disaggregation brings important benefits. \cite{bollerslev2013risk} use frequency-based decomposition to separate the S\&P 500 and the volatility index (VIX) into various frequency components. They find strong coherence between volatility and the volatility-risk reward at low-frequency.

Recently, \cite{bandi2015business} employ frequency-based decomposition for business-cycle consumption risk and asset prices dynamics across horizons using generalized Wold representation (decomposition to wavelet details series). They show the importance of disentangling high- and low-frequency consumption cycles components for pricing of risky assets, and represent the beta of an asset as a linear combination of frequency-specific betas. \cite{dew2016asset} decompose economic fluctuations at different scales and measure the price of risk of consumption fluctuations at each frequency (i.e., frequency-specific risk) discussing asset pricing in the frequency domain. \cite{bidder2016long,dew2017risky} further use frequency domain to study long-run risks.

In addition, \cite{boons2015horizon} argue that horizon-specific macroeconomic risks are key to understanding the link between risk premia and the real economy. \cite{li2017short} study the impact of short-run and long-run consumption risks on the momentum and provide a consumption-based explanation for cross-sectional stock returns. \cite{crouzet2017model} develop a rational expectations model of financial trade with investors who have information available at a range of different frequencies. Finally, \cite{bandi2016economic} disentangle low- and high-frequency components in predictive regression of future excess market returns onto past economic uncertainty. They show that both regression components have scale-specific predictability on low frequency. Other studies using the frequency domain in asset pricing models include \cite{otrok2002habit}, \cite{genccay2005multiscale} and \cite{yu2012using}. 

Wavelet transform plays special role in this literature since a wavelet, being the basic building block of the transform, is a localized filter that is able to work with non-stationary data. As wavelets allow for time-scale decomposition of stochastic processes \citep{antoniou1999}, we allow for the time-scale decomposition in our framework. However, there are drawbacks when the wavelet transform is used. We need to address boundary conditions, as well as stay conscious when building forecasting models due to the use of filters \citep{Gencay2002}. While we are inspired by previous encouraging works using wavelets in precise jump detection and variance estimation \citep{fanwang2008,xue2014jump, barunik}, we explore the possibility of using wavelets in a multivariate setting in order to decompose contributions of continuous and discontinuous parts of covariation.

\section{Estimation of the covariance matrix and co-jumps}\label{est}

To set out the notation, consider the observed $d$-variate (log) price process $(\mathbf{Y}_t)_{t\in[0,T]}$ with $\ell=1,\ldots,d$ components $Y_{t,\ell}$
representing currency prices, i.e., $\mathbf{Y}_t=\left(Y_{t,\ell_1},\ldots,Y_{t,\ell_d}\right)'$. The common assumption regarding the observed prices is that we can decompose the prices into an underlying (log) price process $(\mathbf{X}_t)_{t\in[0,T]}$ and a zero mean
\textit{i.i.d.} noise term $(\boldsymbol \epsilon_t)_{t\in[0,T]}$ with finite
variance that captures microstructure noise. Assuming the noise independent of the price process, we define the observed price process as $\mathbf{Y}_t=\mathbf{X}_t+\boldsymbol \epsilon_t$.

Further, let the $\ell_1$-th and $\ell_2$-th component of the latent process $\mathbf{X}_{t}$ evolve over time as
\begin{eqnarray}
\mathrm{d} X_{t,\ell_1}&=&\mu_{t,\ell_1}\mathrm{d}t+\sigma_{t,\ell_1}\mathrm{d}B_{t,\ell_1} + \mathrm{d} J_{t,\ell_1} \\
\mathrm{d} X_{t,\ell_2}&=&\mu_{t,\ell_2}\mathrm{d}t+\sigma_{t,\ell_2}\mathrm{d}B_{t,\ell_2} + \mathrm{d} J_{t,\ell_2},
\end{eqnarray}
for $\ell_1,\ell_2 \in {1,\ldots,d}$, where $\mu_{t,\ell_i}$ and
$\sigma_{t,\ell_i}$ are c\`adl\`ag stochastic processes,
$B_{t,\ell_i}$ is a standard Brownian motion correlated with
$\rho_t^{\ell_1,\ell_2}=corr(B_{t,\ell_1},B_{t,\ell_2})$, and
$J_{t,\ell_i}$ denotes a (right-continuous) pure jump process for $i=\{1,2\}$. We assume the jump process to have a finite activity, i.e., only a finite number of jumps occur in a finite time interval, and the jump processes can be correlated.

Following standard statistical methods \citep{protter}, the quadratic return covariation associated with $(X_{t,\ell_1},X_{t,\ell_2})$ over the fixed time interval $[0,T]$ can be decomposed into two parts: the continuous part, also called integrated covariance, $IC_{\ell_1,\ell_2}$, and the discontinuous part -- co-jump variation $CJ_{\ell_1,\ell_2}$ as
\begin{equation}
QV_{\ell_1,\ell_2} = \underbrace{\int_0^T \sigma_{t,\ell_1}\sigma_{t,\ell_1}\mathrm{d}\langle B_{\ell_1},B_{\ell_2}\rangle_t}_{IC_{\ell_1,\ell_2}} + \underbrace{\sum_{0\le t \le T} \Delta J_{t,\ell_1} \Delta J_{t,\ell_2}}_{CJ_{\ell_1,\ell_2}}.
\end{equation}
Note that the term $\Delta J_{t,\ell_1} \Delta J_{t,\ell_2}$ is non-zero only if a co-jump occurs, i.e., when both $\Delta J_{t,\ell_1}$ and $\Delta J_{t,\ell_2}$ are non-zero. The quadratic covariation matrix $\bm{QV}$ holding the quadratic variation for $\ell_1=\ell_2$ on the diagonal and quadratic covariation for $\ell_1 \ne \ell_2$ elsewhere can hence be decomposed as
\begin{equation}
\label{qvdecomp}
\bm{QV} = \bm{IC}+\bm{CJ}=\left(
\begin{array}{cc}
IC_{\ell_1,\ell_1} + CJ_{\ell_1,\ell_1} & IC_{\ell_1,\ell_2}+ CJ_{\ell_1,\ell_2}\\
IC_{\ell_2,\ell_1} + CJ_{\ell_2,\ell_1}& IC_{\ell_2,\ell_2}+ CJ_{\ell_2,\ell_2}
\end{array}
\right).
\end{equation}
To study the impact of co-jumps on the dependence structures in currency markets, we are interested in modeling both components of the \autoref{qvdecomp}: the daily ex-post continuous covariation and co-jumps. A usual first step to build the estimator of quadratic covariation is to consider the realized covariance \citep{abdl2003,barndorff2004b} that can be estimated over a fixed time interval $[0\le t\le T]$ as
\begin{equation}
\label{rc}
\widehat{QV}_{\ell_1,\ell_2}^{(RC)}=\sum_{i=1}^N \Delta_i Y_{t,\ell_1} \Delta_i Y_{t,\ell_2},
\end{equation}
where $\Delta_i Y_{t,\ell} = Y_{t+i/N,\ell}-Y_{t+(i-1)/N,\ell}$ is the $i$-th intraday return over the fixed time interval $[0,T]$.

As detailed in \cite{abdl2003} and \cite{barndorff2004b}, realized covariance consistently estimates the quadratic covariation provided that the processes are not contaminated with microstructure noise. Whereas the estimator in \autoref{rc} thus estimates the covariation associated with $(Y_{t,\ell_1},Y_{t,\ell_2})$, we are interested in estimating the covariation associated with $(X_{t,\ell_1},X_{t,\ell_2})$. Several estimators capable of recovering the covariation of the latent process from observed data have been proposed in the literature. A two-scale covariance estimator \citep{Zhang2011} based on subsampling and multivariate kernel-based estimation \citep{BN2011}, which provides a positive semi-definite variance-covariance matrix, are the most notable frameworks. Unfortunately, these approaches can estimate the covariation associated with $(X_{t,\ell_1},X_{t,\ell_2})$ but are not able to decompose it and recover co-jumps. In the following sections, we propose an estimator that will be able to estimate both parts.

\subsection{Co-jump detection}

In order to study the role of co-jumps on correlation structures, we propose simple method for precise localization of co-jumps using the frequency domain tools with special attention to wavelets. In our estimation strategy, we assume that the sample path of the price process has a finite number of jumps (a.s.), i.e., we assume finite jump activity. Building on the theoretical results of \cite{wang95} regarding wavelet jump detection in deterministic functions with \textit{i.i.d.} additive noise, which were recently extended to stochastic processes by \cite{fanwang2008} and \cite{barunik}, we use the discretized version of the continuous wavelet transform to localize co-jumps. Similar to \cite{fanwang2008}, we use the first scale of the discrete wavelet transform to distinguish between the continuous and discontinuous parts of the stochastic price process. The first scale wavelet coefficients represents only the highest frequency, thus they can detect sharp discontinuities in the process without being influenced by other frequency components.

We estimate the co-jump variation associated with $(X_{t,\ell_1},X_{t,\ell_2})$, over $[0\le t\le T]$ in the discrete synchronized time as a sum of co-jumps:
\begin{equation}
\label{cjump}
\widehat{CJ}_{\ell_1,\ell_2} =\sum_{i=1}^N \Delta_i J_{t,\ell_1} \Delta_i J_{t,\ell_2},
\end{equation}
where $\Delta_i J_{t ,\ell}$ is the jump size at intraday position $i$ estimated as
$$\Delta_i J_{t,\ell} =\left(\Delta_i Y_{t,\ell}\right) \mathbbm{1}_{\{ |\mathcal{W}_{1,k}^{\ell}| >\xi \}},$$ where $\mathcal{W}_{1,k}^{\ell}$ denotes the intraday wavelet coefficient at the first scale\footnote{Since we estimate the quadratic covariation on discrete data, we use a non-subsampled version of a discrete wavelet transform, more specifically, the maximal overlap discrete wavelet transform (MODWT). A brief introduction of the discrete wavelet transform and MODWT can be found in \ref{dwt}.}, and $\xi$ is the threshold. As a threshold, we use the universal threshold of \cite{donoho} with the intraday median absolute deviation estimator of standard deviation adapted for the MODWT wavelet coefficients.\footnote{For details, see \cite{PercivalWalden2000}. As we use the MODWT filters, we must slightly correct the position of the wavelet coefficients to obtain the precise jump position; see \cite{PercivalMofjeld1997}.} The threshold $\xi$ has the form: $$\xi=\sqrt{2}\, \text{median}\{|\mathcal{W}_{1,k}^{\ell}|\}\sqrt{2 \log N}/0.6745.$$ If the absolute value of an intraday wavelet coefficient exceeds the threshold $\xi$, then the jump will be estimated at position $k$. In other words, the noise and the continuous part are relatively small, and hence, the dominance of $\mathcal{W}_{1,k}^{\ell}$ results from a discontinuous jump. Then, a co-jump occurs only if both jumps in process $(X_{t,\ell_1},X_{t,\ell_2})$ occur simultaneously. 

In the univariate case, the quadratic jump variation, ${CJ}_{\ell,\ell}$, of the $X_{t,\ell}$ process is estimated as the sum of squares of all of the estimated jump sizes. \cite{fanwang2008} prove that we can estimate the jump variation of the process consistently. Thus, the jump--adjusted price process $Y_{t,\ell}^{(J)}=Y_{t,\ell}-\widehat{CJ}_{\ell,\ell}$ converges in probability to the continuous part without jumps. Because jumps are estimated consistently in $\Delta_i J_{t ,\ell}$ \citep{fanwang2008,barunik}, we can generalize the concept and estimate co-jump variation.

Having estimates of the jump and co-jump variation, the co-jump variation matrix associated with $(X_{t,\ell_1},X_{t,\ell_2})$ can be written as
\begin{equation}
\widehat{\bm{CJ}}=\left(
\begin{array}{cc}
\widehat{CJ}_{\ell_1,\ell_1} & \widehat{CJ}_{\ell_1,\ell_2}\\
\widehat{CJ}_{\ell_2,\ell_1} &\widehat{CJ}_{\ell_2,\ell_2}
\end{array}
\right).
\end{equation}

\subsection{Time-scale decomposition of quadratic covariation}

In this section we show that the quadratic covariation can be decomposed into its frequency components. This result is essential as it allows for construction of our wavelet based integrated covariance estimator.

The quadratic covariation of the discrete process $(Y_{t,\ell_1},Y_{t,\ell_2})$ that belongs to $L^2(\mathbb{R})$ over a fixed time horizon $[0\le t\le T]$ can be expressed as a discrete
wavelet decomposition on a scale-by-scale basis. Hence, for a particular scale $j\in(1,2,\dots)$, we write
\begin{equation}
QV_{\ell_1,\ell_2}(j)=\sum_{k=1}^{N}\mathcal{W}_{j,k}^{\ell_1}  \mathcal{W}_{j,k}^{\ell_2},
\end{equation}
where $\mathcal{W}_{j,k}^{\ell}$ is the intraday wavelet coefficient, with $N$ intraday observations. Asymptotically, as the number of intraday elements goes to infinity ($N\to\infty$), an infinite number of scales $j$ can be used, and the sum of the decomposed quadratic covariation at scales will always be total quadratic covariation (for proof, see \ref{awc})
\begin{equation}
\label{cenergydec}
QV_{\ell_1,\ell_2}=\sum_{j=1}^{\infty} QV_{\ell_1,\ell_2}(j) =\sum_{j=1}^{\infty} \sum_{k=1}^{N}\mathcal{W}_{j,k}^{\ell_1}  \mathcal{W}_{j,k}^{\ell_2}.
\end{equation}
The application of wavelets in Eq. (\ref{cenergydec}) reveals the contributions of particular wavelet scales (frequency bands) to the overall quadratic covariation {\color{red}{}}QV. Thus, we can identify the parts of the frequency spectrum that are essential for this measure. For estimation of Eq. (\ref{cenergydec}) we use the wavelet covariance estimator $\widehat{QV}_{\ell_1,\ell_2}^{(WRC)}$, see appendix \ref{wrc} for details.

\subsection{Data synchronization: Refresh time}
\label{synchrosection}
One important theoretical assumption that we did not mention above is that the data are assumed to be synchronized, meaning that the prices of the assets were collected at the same time stamp. In practice, trading is non-synchronous, delivering fresh prices at irregularly spaced times, which differ across stocks. Research focusing on non-synchronous trading has been an active field of financial econometrics in past years; see, for example, \cite{hayashiyoshida} and \cite{voevlunde}. This practical issue induces bias in the estimators and may be partially responsible for the Epps effect \citep{epps}, a phenomenon of decreasing empirical correlation between the returns of two different stocks with increasing data-sampling frequency.

\cite{sahalia2011} compare various synchronization schemes available in the literature, and find that the estimates do not differ significantly from the estimates obtained using the Refresh Time scheme \cite{BN2011} for the same type of data used here. Thus, we can restrict ourselves to this synchronization scheme. 

Let $N_{t,q}$ be the counting process governing the number of observations in the $q$-th asset up to time $t$, with times of trades $t_{1,q},t_{2,q},\dots$. Following \cite{BN2011}, we define the refresh time, which we use later in our estimator. We present a generalized multivariate version. 

The first refresh time for $t \in [0,1]$ is defined as
\begin{equation}
\tau_1=\max (t_{1,1},\dots,t_{1,q}),
\end{equation}
for $q=1,\dots,d$ assets, and all subsequent refresh times are defined as
\begin{equation}
\tau_{v+1}=\max (t_{1+N_{\tau_v,q},q},\dots,t_{N_{1+\tau_v,q},q}),
\end{equation}
with the resulting Refresh Time sample being of length $N$, whereas $N_{q}$ denotes the number of trades for an individual asset $q$. $\tau_1$ is thus the first time that all assets record prices, whereas $\tau_2$ is the first time that all asset prices are refreshed. In the following analysis, we will set our clock time to ${\tau_v}$ when using the estimators. Specifically, we will consider the $\tau$-th intraday return of the process $Y_{t,\ell}$, $$\Delta_{\tau} Y_{t,\ell} = Y_{t+\tau/N,\ell}-Y_{t+(\tau-1)/N,\ell}.$$ This approach converts the problem into one where the Refreshed Times' sample size $N$ is determined by the degree of non-synchronicity \citep{BN2011}.

\subsection{Jump wavelet covariance estimator\label{sec:jwtscv}}

Using the time-synchronized jump-adjusted price process $(Y_{t,\ell_1}^{(J)},Y_{t,\ell_2}^{(J)})$, we can propose an estimator of the continuous part of quadratic covariation -- the integrated covariance -- $IC_{\ell_1,\ell_2}$, that is robust not only against jumps but also against noise. Furthermore, using wavelet decomposition, we can separate the integrated covariance into $\mathcal{J}^m+1$ scale components representing the integrated covariance at various frequency bands. Our estimator uses the two-scale covariance estimator described by \cite{Zhang2011} and wavelet decomposition. More specifically, we decompose the covariance into wavelet scales $\mathcal{J}^m+1$, and on each scale, we estimate the covariance using the \cite{Zhang2011}'s estimator. Finally, we sum all of the wavelet scales to obtain the final estimate of covariance at all frequencies.

Denote $\widehat{IC}_{\ell_1,\ell_2}^{(JWC)}$ as the jump wavelet estimator (JWC) of the integrated covariance of the asset return processes $(X_{t,\ell_1},X_{t,\ell_2})$ in $L^2(\mathbb{R}$) over the fixed time interval $[0\le t\le T]$. The estimator is defined in terms of the time-synchronized jump-adjusted observed process $(Y_{t,\ell_1}^{(J)},Y_{t,\ell_2}^{(J)})$ as
\begin{equation}
\label{jwtscv}
\widehat{IC}_{\ell_1,\ell_2}^{(JWC)}=\sum_{j=1}^{\mathcal{J}^m+1} c_N\left( \widehat{IC}_{\ell_1,\ell_2}^{(G,J)}(j)-\frac{\bar{n}_G}{n_S} \widehat{IC}_{\ell_1,\ell_2}^{(WRC,J)}(j) \right).
\end{equation}
The estimator consists of two parts: The first part is the averaged version of the estimator (\ref{wrc}) on a grid size of $\bar{n}=N/G$ for a specific wavelet scale $j$:
\begin{equation}
\widehat{IC}_{\ell_1,\ell_2}^{(G,J)}(j)=\frac{1}{G} \sum_{g=1}^G  \sum_{k=1}^{N} \mathcal{W}_{j,k}^{\ell_1}\mathcal{W}_{j,k}^{\ell_2},
\end{equation}
where the wavelet coefficients $\mathcal{W}_{j,k}^{\ell}$ are estimated based on the jump-adjusted process $\Delta Y_{t,\ell}^{(J)} = (\Delta_1 Y_{t,\ell}^{(J)},\ldots,\Delta_N Y_{t,\ell}^{(J)})$. The second term in the estimator (\ref{jwtscv}) denotes the part of the estimator (\ref{wrc}) corresponding to a wavelet scale $j$:
\begin{equation}
\widehat{IC}_{\ell_1,\ell_2}^{(WRC,J)}(j)=\sum_{k=1}^{N}  \mathcal{W}_{j,k}^{\ell_1} \mathcal{W}_{j,k}^{\ell_2}.
\end{equation}
The constant $c_N$ can be tuned for small sample performance, $\bar{n}_G=(N-G+1)/G$, and the same applies for $\bar{n}_S$ (we use $S=1$, and thus, $\bar{n}_S=N$ ). Because both $\widehat{IC}_{\ell_1,\ell_2}^{(G,J)}(j)$ and $\widehat{IC}_{\ell_1,\ell_2}^{(WRC,J)}(j)$ represent the contributions of a specific wavelet scale $j$ only, the final estimator $\widehat{IC}_{\ell_1,\ell_2}^{(JWC)}$ is the sum across all available wavelet scales $j=1,\ldots,\mathcal{J}^m+1$.

Note that the estimator (\ref{jwtscv}) is a sum of the \citep{Zhang2011}'s estimators for all available wavelet scales; hence, the overall speed of convergence of our estimator is governed by the \citep{Zhang2011}'s estimator. Because the \citep{Zhang2011}'s estimator has a rather slow rate of convergence of $N^{-1/6}$ and because the wavelet (variance) covariance estimator converges at rate $N^{-1/2}$ \citep{serroukhwalden2000a}, our estimator converges at a rate of $N^{-1/6}$, and the asymptotic variance is not increased by wavelet decomposition as a result of the variance-preserving property of the wavelets:
\begin{equation}
\widehat{IC}^{(JWC)}_{\ell_1,\ell_2}\overset{p}{\rightarrow}IC_{\ell_1,\ell_2}.
\end{equation}

To estimate the full (variance) covariance matrix $\widehat{\bm{IC}}^{(JWC)}$, we must also estimate the diagonal terms: the integrated variances $\widehat{IC}^{(JWC)}_{\ell_1,\ell_1}$ and $\widehat{IC}^{(JWC)}_{\ell_2,\ell_2}$. These diagonal terms are estimated with our jump wavelet covariance estimator on $Y_{t,\ell_1}^{(J)}$ or $Y_{t,\ell_2}^{(J)}$ separately. This estimation procedure is similar to the jump wavelet two-scale realized variance  estimator for integrated variance proposed by \cite{barunik}. The integrated covariance matrix is
\begin{equation}
\widehat{\bm{IC}}^{(JWC)}=\left(
\begin{array}{cc}
\widehat{IC}^{(JWC)}_{\ell_1,\ell_1} & \widehat{IC}^{(JWC)}_{\ell_1,\ell_2}\\
\widehat{IC}^{(JWC)}_{\ell_2,\ell_1} & \widehat{IC}^{(JWC)}_{\ell_2,\ell_2}
\end{array}
\right).
\end{equation}
With the estimates of the covariance matrix, it is straightforward to compute the correlations from its elements.

\section{Small sample performance of the proposed estimator\label{ch13}}

Before turning our attention to empirical data, we investigate the small sample performance of the jump wavelet covariance estimator to determine how well it can separate the continuous covariation part from quadratic covariation under the presence of noise. In the simulation, we follow the setup of \cite{BN2011} and simulate a bivariate factor stochastic volatility model for $X_{t,i}$, $i=\{1,2\}$ and $t\in[0,1]$ as
\begin{eqnarray}
\label{MCmodel2}
\nonumber dX_{t,i} &= &\mu_{i}dt+\gamma_i \sigma_{t,i} dB_{t,i}+\sqrt{1-\gamma_i^2} \sigma_{t,i} dW_{t}+c_{t,i} dN_{t,i}\\
\nonumber d\sigma_{t,i} &=&\exp(\beta_0+\beta_1 v_{t,i}) \\
dv_{t,i}&=&\alpha v_{t,i} dt+dB_{t,i},
\end{eqnarray}
where the elements of $B_{t,i}$ are independent standard Brownian motions and are independent of $W_t$, and $c_{t,i} dN_{t,i}$ are independent compound Poisson processes with random jump sizes distributed as $N\sim(0,\sigma_{1,J})$. 

The spot correlation between $X_{t,1}$ and $X_{t,2}$ without noise and jumps is used for as a reference: $\sqrt{(1-\gamma_1^2)(1-\gamma_2^2)}$, which is equal to 0.91 here. The full spot covariance matrix
$\Sigma_t=\left(
\begin{array}{cc}
\Sigma_t^{11} & \Sigma_t^{12} \\
\Sigma_t^{12} & \Sigma_t^{22}
\end{array}
\right)=
\left(
\begin{array}{cc}
\sigma_{t,1}^2 &\sigma_{t,1,2} \\
\sigma_{t,1,2} & \sigma_{t,2}^2
\end{array}
\right)$, where $\sigma_{t,1,2}=\sigma_{t,2} \sigma_{t,2} \rho_t$. 

We simulate the processes using the Euler scheme at a time interval of $\delta=1s$, each with $6.5\times60\times60$ steps $n=23,400$, corresponding to a 6.5-trading hour day. The parameters are set to $(\mu_1,\mu_2,\beta_0,\beta_1,\alpha,\gamma_1,\gamma_2)=(0,0,-5/16,1/8,-1/40,-0.3,-0.3)$. Each day is restarted with the initial value of $v_{t,i}$ drawn from a normal distribution $N(0,(-2\alpha)^{-1})$. On each simulated path, we estimate $\widehat{\Sigma}_t$ over $T=1$ day.  The results are computed for samplings of 1 minute, 5 minutes, 30 minutes and 1 hour. 

We repeat the simulations with different levels of noise and different numbers of jumps, assuming the market microstructure noise, $\epsilon_t$, to be normally distributed with different standard deviations: $(E[\epsilon^2])^{1/2}$=\{0,0.0015\}. Thus, we consider simulations with zero noise and 0.15\% of the value of the asset price level noise. We also add different levels of jumps, controlled by intensity $\lambda$ from the Poisson process $c_{t,i} dN_{t,i}$, starting with $\lambda=0$, and continue adding jumps with sizes corresponding to a one standard deviation jump change. We start by simulating prices with only a single co-jump, and then add one jump to each of the bivariate series that are independent of each other.

We use the following benchmark estimators:  the realized covariance (Eq.\ref{rc}), the bipower realized covariance of \cite{barndorff2004}, the two-scale realized covariance of \cite{Zhang2011}, the multivariate realized kernel of \cite{BN2011}, and our jump wavelet covariance estimator (Eq. \ref{jwtscv}).  The realized covariance estimator is neither robust to noise nor can detect co-jumps. The bipower realized covariance is still one of the most popular methods in the literature for the continuous covariance part estimation. This estimator is able to separate the co-jump component of covariation.\footnote{The bipower realized covariance is a natural benchmark for co-jump detection as it is easy to implement and there is no need for calibration of fine tuning parameters.} Moreover, the multivariate realized kernel and two-scale realized covariance are both robust to noise, however they are both unable to separate the continuous and discontinuous (co-jump) part of the quadratic covariation. 

The integrated covariance estimation results are reported in Table \ref{simul_cov1}. Clearly, our estimator can efficiently estimate the continuous covariance of the process in the presence of (co-)jumps and noise. The bipower realized covariance estimator can handle jumps to some extent, whereas as expected, the two-scale realized covariance consistently estimates the quadratic covariation but cannot separate the integrated covariance and co-jumps. This is also the case for the multivariate realized kernel estimator. The sampling frequencies do not reveal any patterns probably because of the effect of quite large jumps in the simulations.

\section{Impact of co-jumps on correlations in currency markets\label{ch34}}

The primary aim of this work is to shed light on the sources of dependence in currency markets, especially relating to the role of co-jumps. The proposed methodology is an efficient way of estimating both parts of quadratic covariation, and thus, we use it to determine total and continuous correlation between currency pairs. In addition, we study the roles of the different trading sessions during the day. 

\subsection{Data description}

We study the relationship among the British pound (GBP), Swiss franc (CHF) and euro (EUR) futures logarithmic price returns, specifically the GBP--CHF, GBP--EUR and CHF--EUR currency pairs. Currency future contracts are traded on the Chicago Mercantile Exchange (CME) and are quoted in the unit value of the foreign currency in US dollars, which makes them comparable. It is advantageous to use currency futures data for this analysis instead of spot currency prices because the former embed interest rate differentials and do not suffer from additional microstructure noise from over-the-counter trading. The cleaned data are available from Tick Data, Inc.\footnote{http://www.tickdata.com/}

It is important to understand the trading system before we proceed with the estimation. In August 2003, CME launched the Globex trading platform, which substantially increased the liquidity of currency futures. On Monday, December 18, 2006, the CME Globex$^{\tiny \textregistered}$ electronic platform started offering nearly continuous 23-hour-a-day trading. The weekly trading cycle begins at 17:00 Central Standard Time (CST) on Sunday and ends at 16:00 CST on Friday. Each day, the trading is interrupted for one hour from 16:00 CST until 17:00 CST. These changes in the trading system dramatically affected trading activity. For this reason, we restrict ourselves to a sample period extending from January 5, 2007 through July 3, 2015, which includes the recent financial crisis. The futures contracts we use are automatically rolled over to provide continuous price records, and thus, we do not have to address different maturities.

We divide the 23-hour trading day into three trading sessions: Asia (17:00 -- 2:00 CST), which lasts for 9 hours; Europe (2:00 -- 8:00 CST), which lasts for 6 hours; and the U.S. (8:00 -- 16:00 CST), which lasts for 8 hours. We exclude potential jumps resulting from the one-hour gap in trading from our analysis by redefining the day in accordance with the electronic trading system. Moreover, we eliminate Saturdays and Sundays, US federal holidays, December 24 to December 26, and December 31 to January 2, because of the very low activity on these days, which would bias the estimates.

Looking more closely at the higher frequencies, we find that many transactions have a common time stamp. For these occasions, we use arithmetic average values for all observations with the same time stamp. Finally, we redefine the clock according to the refresh time scheme to obtain synchronized data. We use the refresh time scheme (Section \ref{synchrosection}) for each pair separately to retain as much data as possible in the analysis. Table \ref{tabdesc} displays the descriptive statistics for the logarithmic futures returns with frequencies of 1 minute and 5 minutes.

\begin{figure}[t]
\centering
\includegraphics[width=\textwidth]{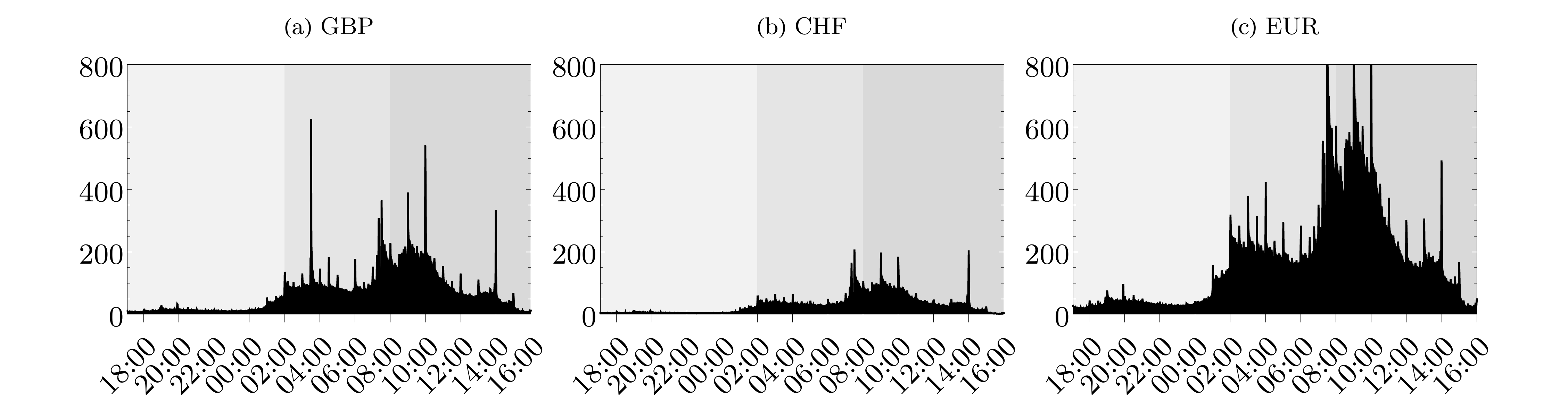}
\caption{Trading activity on (a) GBP, (b) CHF, and (c) EUR future contracts measured in terms of the average volume using 1-minute trading intervals over the whole period of January 5, 2007 -- July 3, 2015. The trading session hours from Asia (17:00 -- 2:00 CST) to Europe (2:00 -- 8:00 CST) and then to the U.S. (8:00 -- 16:00 CST) are highlighted by different background shades.}
\label{volume}
\end{figure}

Next, it is important to see trading activity of the three currency futures in the different sessions during the day. We measure the trading activity of the futures using 1-minute intervals. For a given minute, we compute the average over the whole sample and thus obtain a clear picture of how trading activity on FX markets is distributed. Figure \ref{volume} shows low volumes in Asia relative to the sessions in Europe and the U.S. Trading activity peaks before the most active U.S. session starts. When we examine trading activity in terms of currencies, CHF displays the lowest volume, followed by GBP, whereas the most actively traded currency in our selection is EUR.

\subsection{Exact co-jump detection}

Since the distribution of the estimated jump and co-jump variation is unknown, a testing strategy using bootstrapping is appropriate. In addition, bootstrapping can significantly improve the finite sample properties of the jump \citep{dovonon2014bootstrapping} and co-jump tests based on realized measures. Our newly proposed estimator can separate the continuous part of covariance from processes that include co-jumps and are contaminated with noise. If we were interested in actually estimating the co-jumps from the observed data, we could compare them with the quadratic covariation estimate, and considering the estimation error of both estimators, a standard Hausman-type test statistic could be proposed. In a univariate setting, \cite{barunikjwg} bootstrap this type of statistic for jump detection. Here, we extend the univariate approach and use a bootstrap testing procedure to test for the presence of jumps and co-jumps in a given time interval. From now on, we will be using only the bootstrapped version of our jump wavelet estimator denoted with asterix. Technical details are given in \ref{sec:bootstrap}.

Being able to identify days that have significant co-jump components, next step is to determine whether the reason why the null hypothesis of no co-jumps was rejected is because of the presence of co-jump(s) or, alternatively, because of the occurrence of large idiosyncratic (disjoint) jump(s). \cite{gnabo2014system} show that large idiosyncratic jumps may inflate the test statistic, and thus, co-jumps may be falsely detected. Therefore, there are basically two possible reasons why the null hypothesis was rejected:
\begin{enumerate}
\item Co-jumps: $t\in[0,T]$: $\Delta_i J_{t,\ell_1} \Delta_i J_{t,\ell_2} \neq 0$, i.e., the process is not exactly zero.
\item Disjoint jumps: $t\in[0,T]$: the processes $\Delta_i J_{t,\ell_1}$ and $\Delta_i J_{t,\ell_2}$ are not both zero (at least one of them), but $\Delta_i J_{t,\ell_1} \Delta_i J_{t,\ell_2} = 0$.
\end{enumerate}

An advantage of our approach is that the exact jump position is obtained by the wavelet analysis; hence, we can successfully eliminate the false co-jump situation caused by high idiosyncratic jump(s). Furthermore, because we know the directions of the jumps, we can distinguish between co-jumps that occur with jumps of the same or different direction on day $t$.

\subsection{Covariance}

\begin{figure}[!h]
\centering
\includegraphics[width=\textwidth]{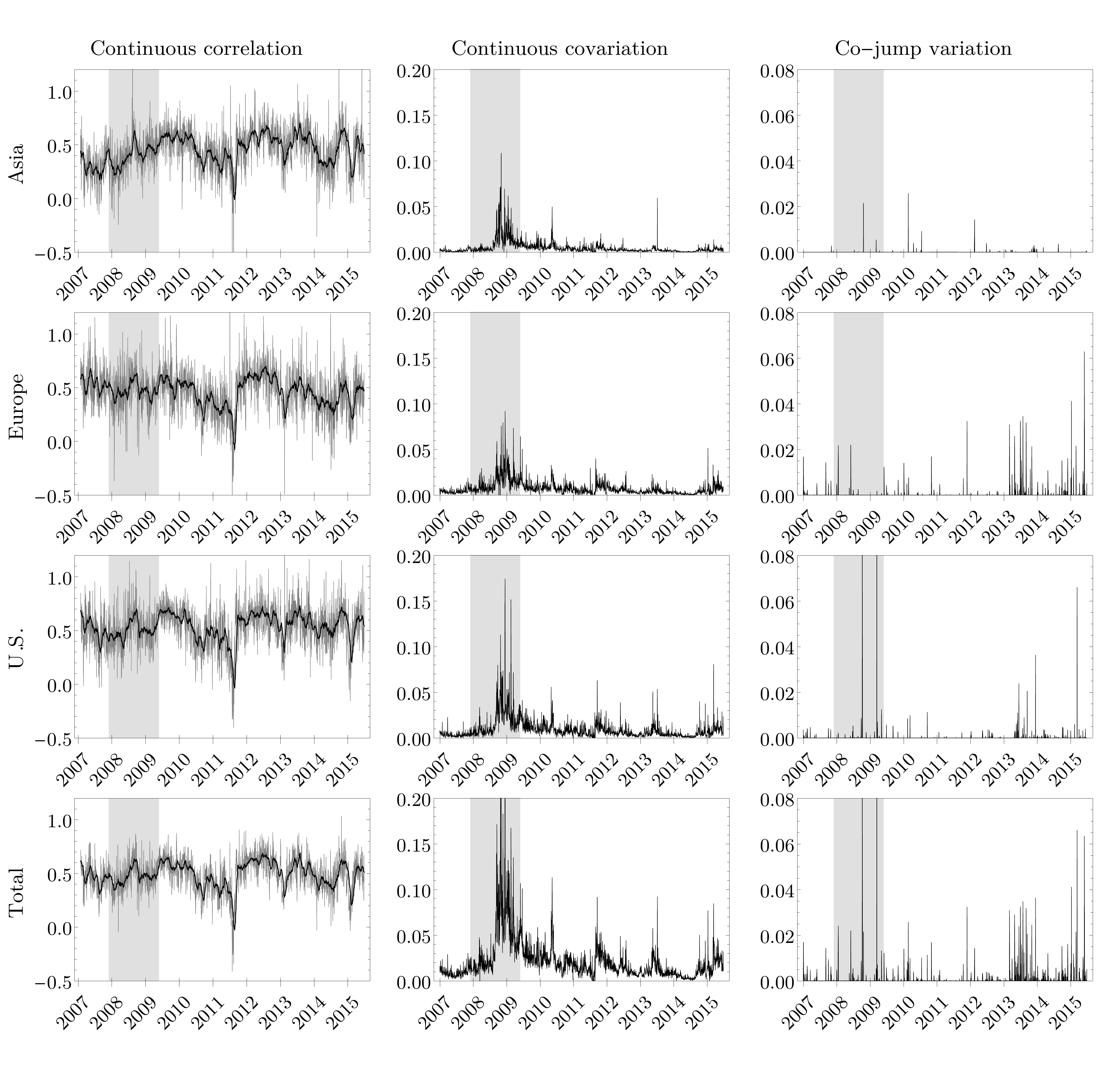}
\caption{\textbf{GBP--CHF pair}: Continuous correlation in gray with a 21-day moving average in black (left column), integrated covariance (middle column), and co-jumps (right column) estimated by jump wavelet covariance estimator. The quantities computed during the Asian, European, and U.S. sessions are depicted in the first three rows. The last row lists the quantities computed over a whole trading day session. The 2007 -- 2008 crisis period is shaded.}
\label{gbpchf}
\end{figure}

\begin{figure}[!h]
\centering
\includegraphics[width=\textwidth]{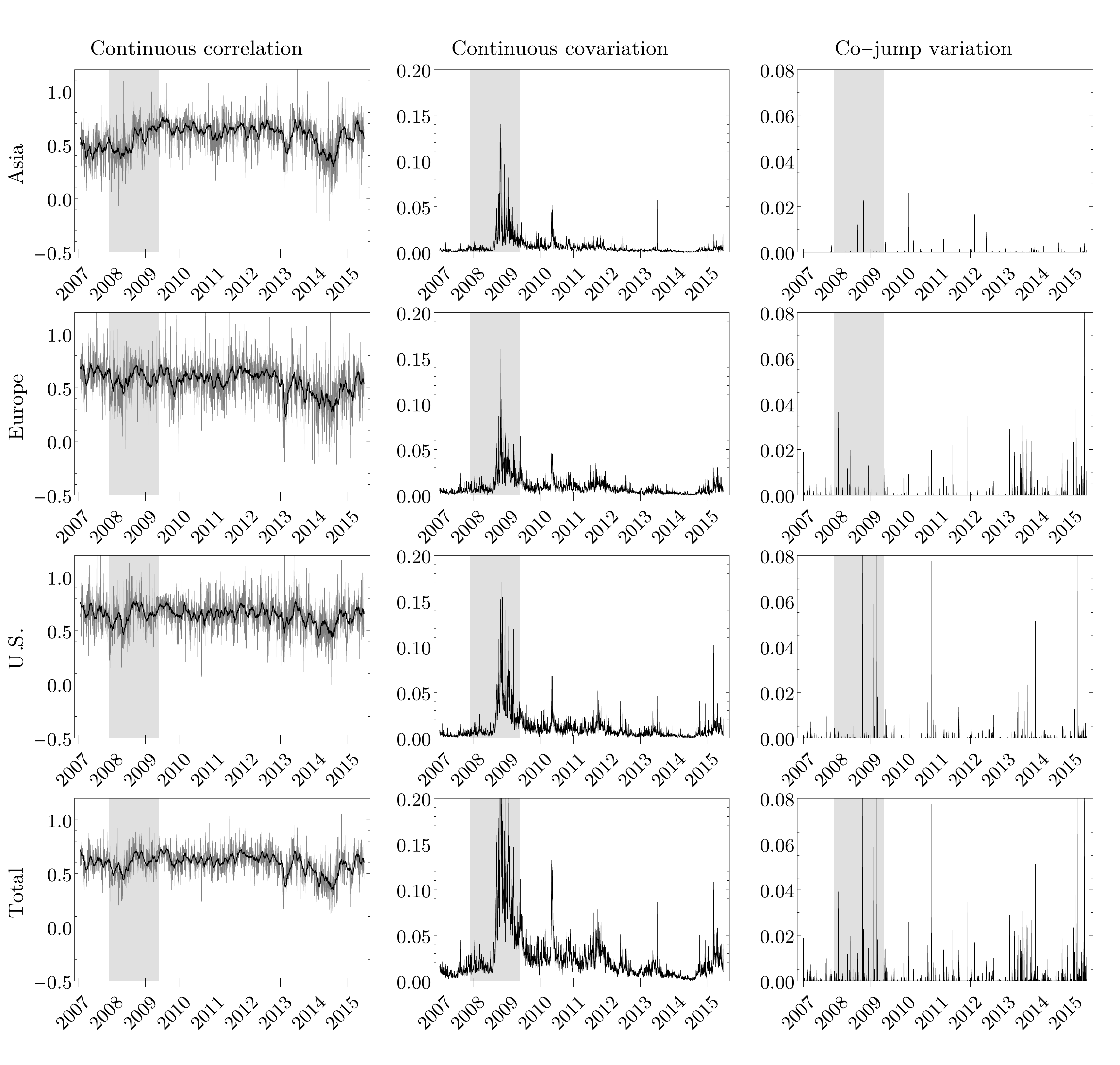}
\caption{\textbf{GBP--EUR pair}: Continuous correlation in gray with a 21-day moving average in black (left column), integrated covariance (middle column), and co-jumps (right column) estimated by jump wavelet covariance estimator. The quantities computed during the Asian, European, and U.S. sessions are depicted in the first three rows. The last row lists the quantities computed over a whole trading day session. The 2007 -- 2008 crisis period is shaded.}
\label{gbpeur}
\end{figure}

\begin{figure}[!h]
\centering
\includegraphics[width=\textwidth]{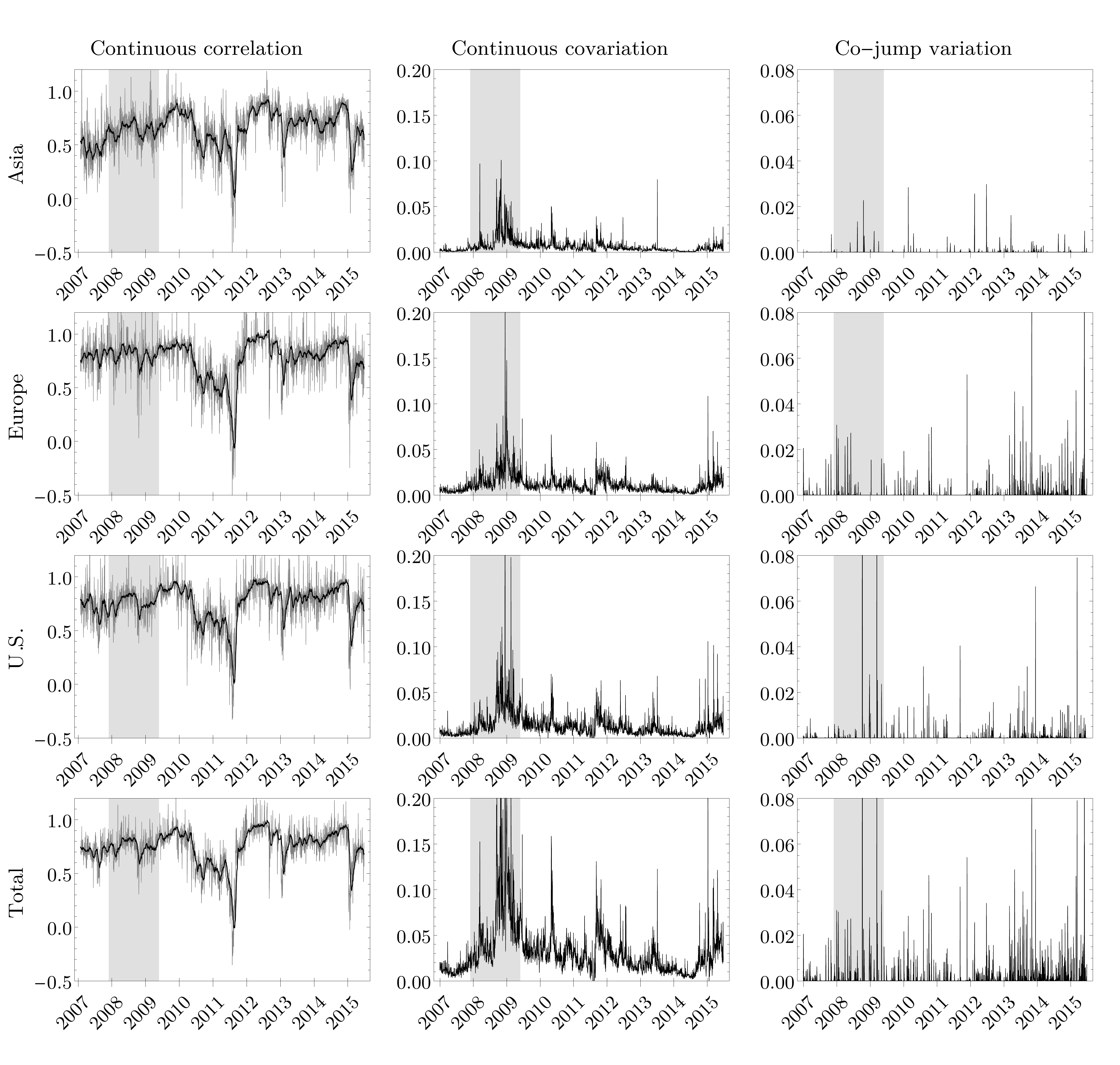}
\caption{\textbf{CHF--EUR pair}: Continuous correlation in gray with a 21-day moving average in black (left column), integrated covariance (middle column), and co-jumps (right column) estimated by jump wavelet covariance estimator. The quantities computed during the Asian, European, and U.S. sessions are depicted in the first three rows. The last row lists the quantities computed over a whole trading day session. The 2007 -- 2008 crisis period is shaded.}
\label{chfeur}
\end{figure}

We estimate the covariance matrix of the three currency pairs GBP--CHF, GBP--EUR and CHF--EUR using the newly proposed jump wavelet covariance estimator. The middle column of Figures \ref{gbpchf}--\ref{chfeur} show the estimates of continuous covariation for different trading sessions. The evolution of the covariance over time reveals that all pairs were exposed to increased covariance during the financial crisis of 2007--08 (highlighted in gray in Figures \ref{gbpchf}--\ref{chfeur}). Furthermore, increased activity for the CHF--EUR pair can be observed in 2015, which may be partially caused by the strong appreciation of the CHF after the surprising decision of the Swiss national bank to remove its cap on the CHF on January 15, 2015.\footnote{The CHF soared more than 30\% percent relative to the Euro on January 15, 2015.} Table \ref{tabjnum} further summarizes the results. The highest covariance is measured for the CHF--EUR pair, whereas the GBP--CHF pair shows the lowest values. Analogously to the trading activity discussed in the previous paragraphs, we observe the lowest covariance in the Asian trading session and the highest in the U.S. session. 

\begin{table}[!h]
\caption{Number of days with co-jumps, co-jump variation distribution among trading sessions (CJ-d), quadratic covariation (QV) and the ratio of co-jumps variation to quadratic covariation (\% CJ/QV), maximum values are shown in bold.}
\small
\begin{center}
\begin{tabular}{l rccccccc}
\toprule
\multicolumn{2}{c}{} & \multicolumn{2}{c}{Asia}& \multicolumn{2}{c}{EU}& \multicolumn{2}{c}{U.S.}& Total \\
\cmidrule(r){3-9}
&& \# &\% & \# &\% & \# &\%& \\
\midrule
\multirow{4}{*}{GBP--CHF}
&Days with CJ$\neq$ 0& 57&17.0 & 139 &\textbf{41.5} &139&\textbf{41.5}&335  \\
&CJ-d& - & 7.4& - &\textbf{47.7} & - &44.8& -  \\
&QV&0.046&21.8&0.073& 34.6 &0.093&\textbf{44}&0.211\\
&\% CJ/QV& - & 0.6 & - & \textbf{2.4} & - & 1.5&- \\
 \cmidrule(r){2-9}
\multirow{4}{*}{GBP--EUR}
&Days with CJ$\neq$ 0& 82&18.9 & 156 &36.1&194&\textbf{44.9}&432  \\
&CJ-d& - &7.4 & - & 42.4 & - & \textbf{50.0}&-\\
&QV& 0.061 &23.6&0.087& 33.8 &0.110&\textbf{42.8}& 0.257\\
&\% CJ/QV& - & 0.82 & - & \textbf{3.0}& - & 2.0&- \\
 \cmidrule(r){2-9}
\multirow{4}{*}{ CHF--EUR}
&Days with CJ$\neq$ 0& 122&18.9 & 246 &38.3 &275&\textbf{41.7}&643  \\
&CJ-d&-&8.8 &-&\textbf{46.7}  &-& 44.4 &-\\
&QV& 0.066 &20.9 &0.114& 36.2 &0.136&\textbf{43.2}&0.315\\
&\% CJ/QV& - & 1.2 & - & \textbf{3.6} & - & 2.6&- \\ 
\bottomrule
\end{tabular}
\end{center}
\label{tabjnum}
\end{table}

\begin{table}
\caption{Unconditional correlations measured during the Asia, EU, and U.S. trading hours.}
\small
\begin{center}
\begin{tabular}{l cccc}
\toprule
 & Asia  & EU  & U.S. & Total\\
\midrule
GBP--CHF  &  0.446 & 0.464 & 0.537 & 0.489\\
GBP--EUR  & 0.579 & 0.561 & 0.646 & 0.598\\
CHF--EUR  & 0.646 & 0.771 & 0.758 & 0.745\\
\bottomrule
\end{tabular}
\end{center}
\label{tabcorr}
\end{table}

\subsection{Co-jump variation}

A question we would like to address relates to the importance of co-jumps for the currency pairs and how they impact the covariance and correlation. Before quantifying these effects, we must examine the dynamics of co-jumps themselves.

The right columns of Figures \ref{gbpchf}--\ref{chfeur} reveal that number of days with co-jumps is very low in Asia relative to the EU and U.S. sessions. In addition, Table \ref{tabjnum} shows that less than 20\% of the days with co-jumps occur during the Asian session. This may be attributed to the relatively low trading volumes of the currency pairs in Asia and the minimum of important news reported when the Asian markets are open. In contrast, the EU and the U.S. sessions exhibit similar proportions of days with co-jumps, indicating that news influencing the currency pairs is nearly equally distributed across these markets. We note that the threshold in the co-jump estimation (Eq. \ref{cjump}) is computed separately for all sessions. In this way, we use session-specific thresholding, considering the large differences in variances of prices during the trading day.

\begin{figure}[t]
\centering
\includegraphics[width=\textwidth]{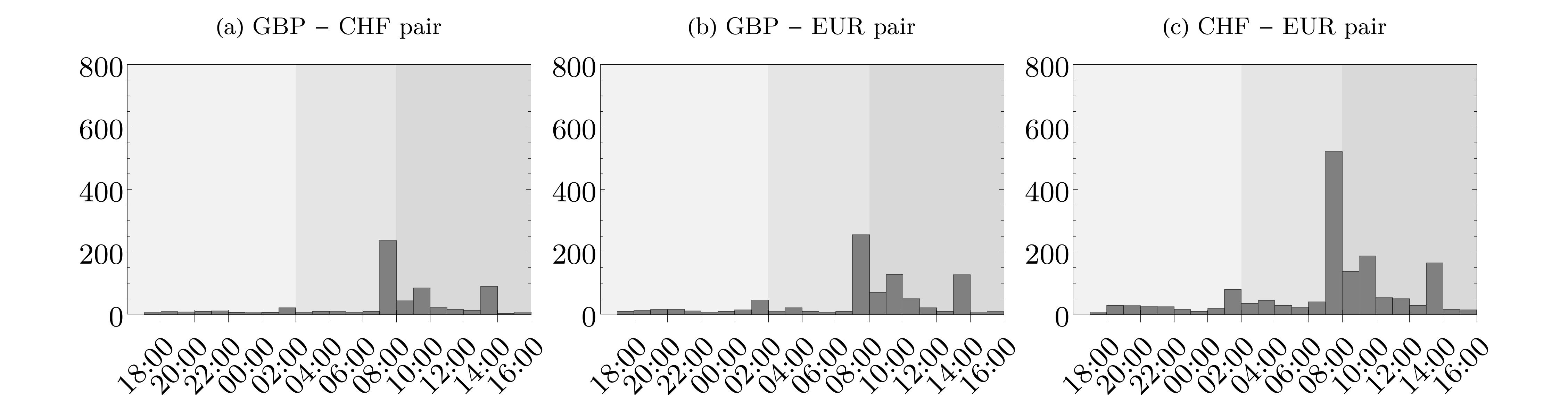}
\caption{Distribution of co-jumps during trading sessions starting with Asia (17:00 -- 2:00 CST), then Europe (2:00 -- 8:00 CST), and then the U.S. (8:00 -- 16:00 CST) highlighted using different background shades.}
\label{ch}
\end{figure}

The magnitude of co-jump variation generally differs across trading sessions. We document high co-jump variation during the EU and the U.S. sessions (Figures \ref{gbpchf}--\ref{chfeur}, Table \ref{tabjnum}). In addition, Figure \ref{ch} documents the number of co-jumps found during the different trading hours. The highest number of co-jumps is generally detected during the U.S. trading session, with its peak one hour before the U.S. trading session starts (7:00--8:00 CST). Interestingly, the largest number of co-jumps is found during the period of low rate of news influencing European currencies. We attribute this finding to the highest trading activity of the futures contracts.

\begin{figure}[h!]
\centering
\includegraphics[width=\textwidth]{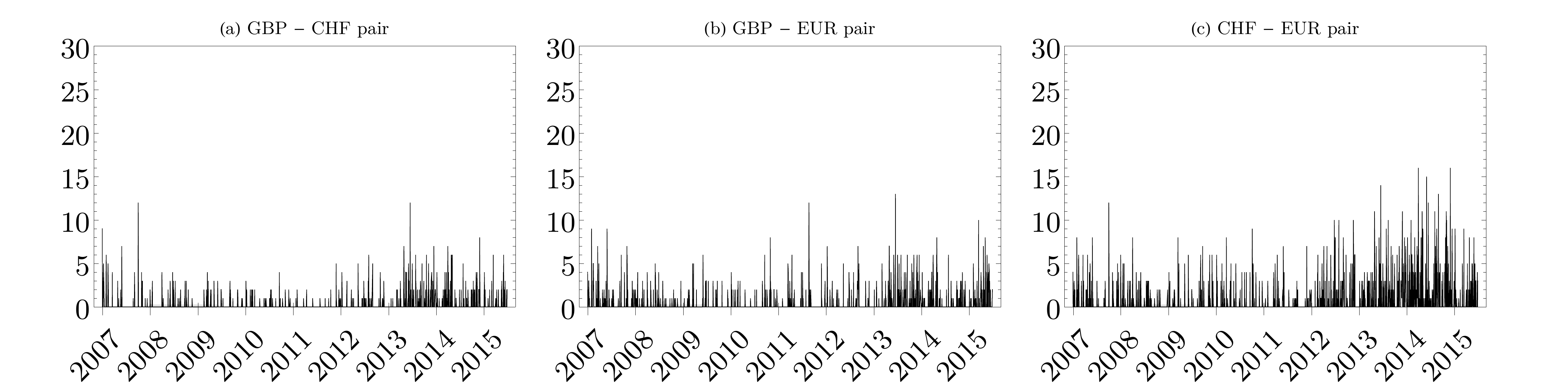}
\caption{Number of co-jumps for all three currency pairs during 2007 -- 2015.}
\label{cm}
\end{figure}

The news influencing the U.S. session perceives the European currency markets as single market, and thus, the differences between GBP, CHF and EUR are small from the U.S. perspective. Another important factor that influences the U.S. session is arbitrage. Because all of the currencies are denominated in U.S. dollars, large shifts in the USD cause subsequent co-jumps for all other currencies.

The situation is very different in the Asian session, where we observe the lowest number of co-jumps for all three currency pairs (see Figure \ref{ch}) and the lowest co-jump variation of less than 10\%. This low co-jump variation corresponds to the low covariance, with only approximately 20\% of the total covariation contributed by the Asian session.

\subsubsection{Co-jump variation dynamics over time}

The results above suggest that the share of co-jump variation differs across trading hours. Another relevant question is whether the proportion of co-jumps, representing the discontinuous part of the quadratic covariation, is stable over time. To observe these dynamics, we divide the sample into years and compute the shares of co-jumps in the quadratic covariation corresponding to a given year (see Table \ref{tabcjd}). The results indicate that the share of co-jump variation in quadratic covariation increased substantially in 2013 and 2014 for all pairs and all sessions (see Figure \ref{cm}). This shows growing importance of co-jumps, thus the accurate detection of discontinuous components is essential. For example, the CHF--EUR pair during the EU session exhibited the highest share of co-jumps in the whole examined period, more specifically, in 2014, this share accounted for more than 9\%, a significant proportion. 

\begin{table}
\small
\caption{Dynamics of the ratio of discontinuous part (co-jumps variation) to quadratic covariation (maximum values are shown in bold)}
\begin{center}
\begin{tabular}{lrrrrrrrrrr}
\toprule
 & & 2007& 2008& 2009 & 2010 & 2011 & 2012 & 2013 & 2014 & 2015 \\
  \cmidrule(r){2-11}
 \multirow{4}{*}{GBP--CHF}
 &\text{Asia} & 0.18 & 0.27 & 0.14 & 1.00 & 0.05 & 0.64 & \textbf{2.10} & 0.72 & 0.10 \\
 &\text{EU} & 2.00 & 1.60 & 0.55 & 1.10 & 1.30 & 0.40 & \textbf{6.20} & 5.60 & 4.00 \\
 &\text{U.S.} & 2.40 & 0.96 & 1.20 & 0.72 & 0.20 & 1.40 & 2.50 & \textbf{2.60} & 1.70 \\
 &\text{Total} & 2.20 & 1.80 & 1.20 & 1.40 & 0.87 & 1.10 & \textbf{6.01} & 6.00 & 3.80 \\
 \cmidrule(r){2-11}
 \multirow{4}{*}{GBP--EUR}
 &\text{Asia} & 0.17 & 0.49 & 0.24 & 0.74 & 0.25 & 1.30 & \textbf{2.00} & 1.00 & 1.50 \\
 &\text{EU} & 2.80 & 2.00 & 2.70 & 0.98 & 1.60 & 1.00 & \textbf{6.10} & 6.00 & 5.00 \\
 &\text{U.S.} & 3.00 & 0.81 & 2.00 & 2.00 & 1.20 & 1.50 & 2.70 & 2.20 & \textbf{3.30} \\
 &\text{Total} & 3.10 & 1.90 & 1.80 & 2.20 & 1.50 & 1.80 & \textbf{5.90} & 4.90 & 5.60 \\
 \cmidrule(r){2-11}
 \multirow{4}{*}{CHF--EUR}
  &\text{Asia} & 0.52 & 0.71 & 0.35 & 0.83 & 0.81 & 1.50 & \textbf{3.60} & 1.90 & 0.76 \\
 &\text{EU} & 2.80 & 1.90 & 0.95 & 2.20 & 1.00 & 2.90 & 7.50 & \textbf{9.30} & 5.00 \\
 &\text{U.S.} & 2.20 & 1.20 & 1.80 & 2.10 & 1.10 & 2.50 & 4.20 & \textbf{5.70} & 2.80 \\
 &\text{Total} & 2.90 & 2.10 & 1.70 & 2.40 & 1.60 & 3.10 & 7.40 & \textbf{9.30} & 4.60 \\
\bottomrule
\end{tabular}
\end{center}
\label{tabcjd}
\end{table}

\subsection{Correlation}

Armed with the precise decomposition of the continuous and discontinuous part of quadratic covariation, we can proceed to our main result and study how co-jumps impact correlations. First, it is useful to look at the total correlation defined as
\begin{equation}
\label{eq:corr}
corr_T^{(t)}=\frac{QV_{\ell_1,\ell_2}}{\sqrt{QV_{\ell_1,\ell_1}}\sqrt{QV_{\ell_2,\ell_2}}}=
\frac{IC_{\ell_1,\ell_2}+CJ_{\ell_1,\ell_2}}{\sqrt{IC_{\ell_1,\ell_1}+CJ_{\ell_1,\ell_1}}\sqrt{IC_{\ell_2,\ell_2}+CJ_{\ell_2,\ell_2}}},
\end{equation}
with quadratic covariation of $(Y_{t,\ell_1},Y_{t,\ell_2})$ being normalized by volatilities of $Y_{t,\ell,\ell}$ processes. Quadratic covariation has two components, and we are mainly interested in studying the influence of co-jump part on correlation structure. Naturally, non-zero idiosyncratic jumps $CJ_{\ell,\ell}$ coming from individual assets will decrease the total correlation, while the presence of co-jumps $CJ_{\ell_1,\ell_2}$ will cause an increase in total correlation. 

Since we want to control for the effects of microstructure noise, the estimators we use in testing are \cite{Zhang2011}'s two-scale realized covariance estimator (TSCV) and our jump wavelet covariance estimator (JWC*). The total correlation is estimated as
\begin{equation}
\label{eq:corrqv}
\widehat{corr}_T^{(t)} = \frac{\widehat{QV}_{\ell_1,\ell_2}^{(TSCV)}}{\sqrt{\widehat{QV}^{(TSCV)}_{\ell_1,\ell_1}}\sqrt{\widehat{QV}^{(TSCV)}_{\ell_2,\ell_2}}}. 
\end{equation}

The continuous correlation, containing only continuous components, thus having neither jumps nor co-jumps, denoted as $corr_T^{(c)}$, is estimated as
\begin{equation}
\widehat{corr}_T^{(c)} =  \frac{\widehat{IC}_{\ell_1,\ell_2}^{(JWC*)}}{\sqrt{\widehat{IC}^{(JWC*)}_{\ell_1,\ell_1}}\sqrt{\widehat{IC}^{(JWC*)}_{\ell_2,\ell_2}}}.
\end{equation}

Let us look at the correlations across the sessions and in time. The left column of Figures \ref{gbpchf}--\ref{chfeur} show the dynamic continuous correlations, and Table \ref{tabcorr} summarizes the unconditional correlation across sessions. We observe generally lower correlations during the Asian session, and higher correlations during the U.S. session. The CHF--EUR exhibits the highest correlation, whereas GBP--CHF has the lowest one. This difference is substantial, exceeding 0.25. The CHF-EUR pair exhibits the richest dynamics of continuous correlations, including two clear periods of very low correlations (mid-2011 and the beginning of 2015), approaching zero.

\subsubsection{How co-jumps impact correlations?}

\begin{figure}[!h]
\centering
\includegraphics[width=\textwidth]{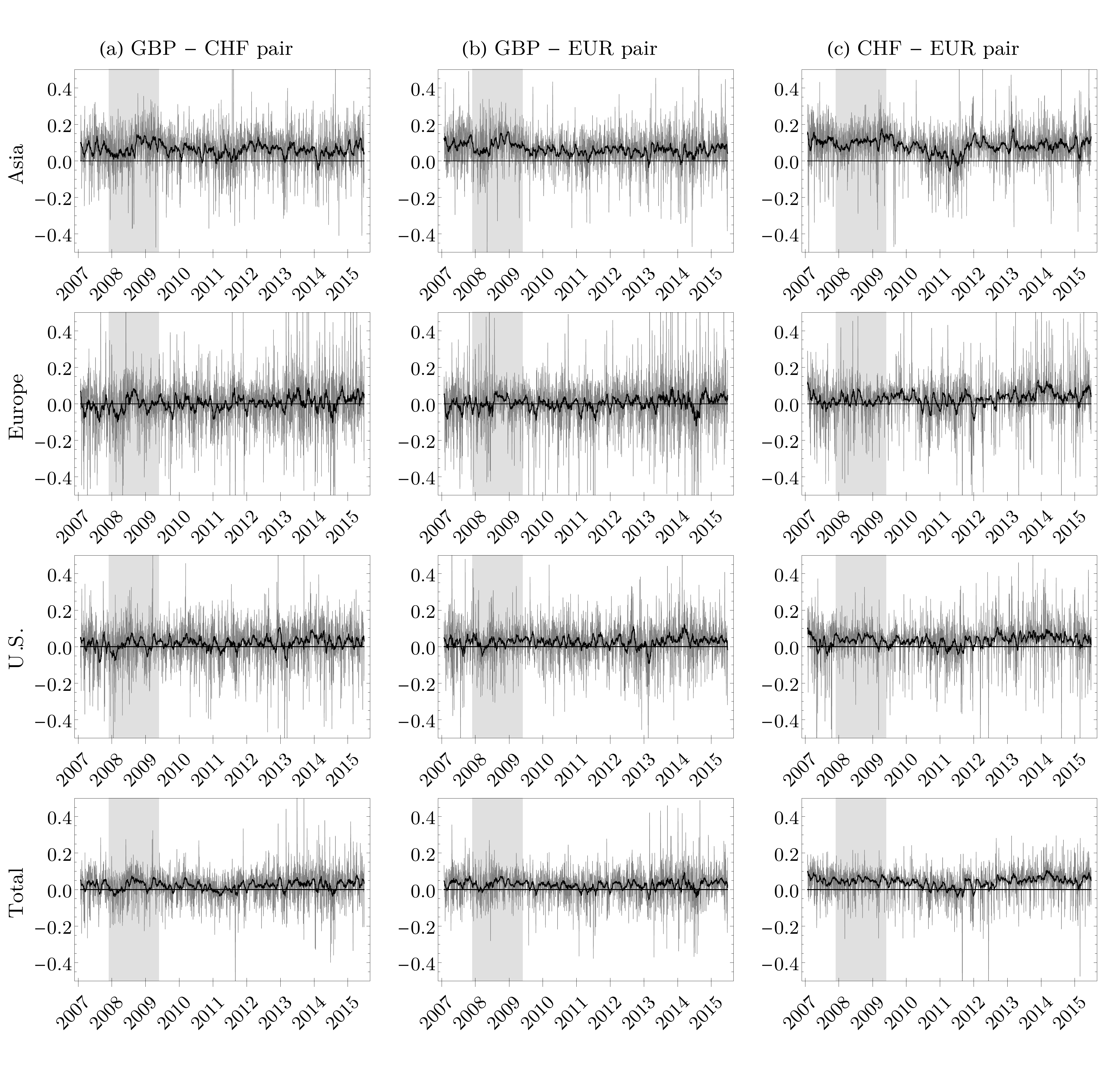}
\caption{Correlation difference $corr_T^{(t)} - corr_T^{(c)}$ in gray with a 21-day moving average in black for the GBP-CHF (left column), GBP-EUR (middle column), and CHF-EUR pairs (right column). The quantities computed during the Asian, European, and U.S. sessions are depicted in the first three rows. The last row lists the quantities computed over a whole trading day session. The 2007 -- 2008 crisis period is shaded.}
\label{corrdiff}
\end{figure}

Since we are able to precisely estimate the jump and co-jumps components, we can study how co-jumps influence correlations. As a first step, we compare the correlation difference, $\widehat{corr}_T^{(t)}-\widehat{corr}_T^{(c)}$, in time for all three pairs and across all trading hours. Figure \ref{corrdiff} shows the difference together with its moving average, medians of these differences are summarized in Table \ref{corrdifftab}. In the event that the correlation difference is positive, i.e., $corr_T^{(t)}\ge corr_T^{(c)}$, the co-jumps are significant part of the total correlation. In other words, the continuous correlation, without the co-jumps and jumps, is lower than the correlation estimated with quadratic covariation and variance estimators. The correlation difference is the highest for the CHF--EUR pair and generally in the Asian session.

\begin{table}[h!]
\caption{Medians of differences between total and continuous correlations measured during the Asia, EU, and U.S. trading hours.}
\small
\begin{center}
\begin{tabular}{l cccc}
\toprule
                  &  Asia  &  EU  &  U.S.  &  Total\\
\midrule
GBP--CHF &0.065 &0.010 &0.030 &0.030\\
GBP--EUR &0.064 &0.013 &0.032 &0.031\\
CHF--EUR &0.087 &0.032 &0.039 &0.052\\
\bottomrule
\end{tabular}
\end{center}
\label{corrdifftab}
\end{table}

Additionally, we can build a simple testing strategy to see whether the correlation differences are statistically significant. Under the null hypothesis of zero impact of jumps and co-jumps on total correlation, the difference between the total and continuous correlation will be zero, as implied by \autoref{eq:corr}. To test the null hypothesis $\mathcal{H}_0: \widehat{corr}_T^{(t)}-\widehat{corr}_T^{(c)} = 0$, we estimate a simple regression
\begin{equation}
\label{eqmz}
\widehat{corr}_T^{(t)}= \alpha + \beta \widehat{corr}_T^{(c)} + \epsilon_T,
\end{equation}
with zero mean \textit{i.i.d.} error with constant variance. In case $\alpha=0$, and $\beta=1$ jointly, we are not able to reject equality of correlations with and without (co-)jumps. Hence, the null hypothesis translates to testing that $\mathcal{H}_0: \alpha = 0 \cap \beta = 1$ against $\mathcal{H}_A: \alpha \ne 0 \cap \beta \ne 1$. Furthermore, we pay special attention to coefficient $\alpha$ since positive $\alpha$ would directly imply that occurrence of co-jumps play important role in total correlation. Conversely, negative alpha would imply that idiosyncratic (individual) jumps have a larger impact on total correlations than co-jumps. Table \ref{tabcojimpact} shows the estimation results and reveals that $\widehat{corr}_T^{(t)}\ge \widehat{corr}_T^{(c)}$, and co-jumps seem to be a significant part of total correlations. The largest impact is seen in the GBP-EUR pair. In terms of sessions, co-jumps seem to have a similar impact on total correlations during all sessions. In all cases, we reject the joint null hypothesis about coefficients using Wald test with heteroscedasticity consistent with White's covariance estimator.

To increase the power of the test, we run additional regressions to \autoref{eqmz}  including instruments such as a lagged variables proxy in the regression to confirm that the results are robust to possible dependence structures in the data, such as nonlinearities and persistence. In addition, we run transformed regression using generalized least squares to control heteroscedasticity, and possibly auto-covariance structures in the residuals of original regression that could impact the size and power of the test. Following \cite{patton2009evaluating}, we estimate parameters using Generalized Least Squares (GLS) as $\widehat{corr}_T^{(t)}/\widehat{corr}_T^{(c)}= \alpha/\widehat{corr}_T^{(c)} + \beta  + \epsilon_T$ regression. All the additional tests\footnote{We do not repeat the results here since they do not add any additional information. Instead, the results are available upon request from authors.} decisively support the previous results with slightly better precision; hence, we can conclude that we document the impact of co-jumps on total correlations.

\begin{table}[!h]
\caption{Impact of co-jumps. Table \ref{tabcojimpact} shows estimated coefficients from the regression $\widehat{corr}_T^{(t)}= \alpha + \beta \widehat{corr}_T^{(c)} + \epsilon_T,$. All cases when $\mathcal{H}_0: \alpha = 0 \cap \beta = 1$ is rejected using Wald test with heteroskedasticity consistent White's covariance estimator are indicated by bold.}
\small
\begin{center}
\begin{tabular}{lrcccc}
\toprule
\multicolumn{2}{c}{} & Asia& EU& U.S.& Total \\
\cmidrule(r){2-6}
\multirow{4}{*}{GBP--CHF}
&$\alpha$& \textbf{0.120} & \textbf{0.093} & \textbf{0.119} & \textbf{0.048} \\
&$\beta$& \textbf{0.868} & \textbf{0.803} & \textbf{0.813} & \textbf{0.947} \\
&$R^2$& 0.713 & 0.581 & 0.665 & 0.746 \\
 \cmidrule(r){2-6}
\multirow{4}{*}{GBP--EUR}
&$\alpha$& \textbf{0.208} & \textbf{0.165} & \textbf{0.261} & \textbf{0.140} \\
&$\beta$& \textbf{0.750} & \textbf{0.711} & \textbf{0.637} & \textbf{0.808} \\
&$R^2$&  0.611 & 0.492 & 0.495 & 0.593 \\
 \cmidrule(r){2-6}
\multirow{4}{*}{ CHF--EUR}
&$\alpha$&  \textbf{0.173} & \textbf{0.168} & \textbf{0.155} & \textbf{0.081} \\
&$\beta$& \textbf{0.861} & \textbf{0.821} & \textbf{0.840} & \textbf{0.948} \\
&$R^2$&  0.751 & 0.722 & 0.733 & 0.833 \\
\bottomrule
\end{tabular}
\end{center}
\label{tabcojimpact}
\end{table}

Impact of co-jumps on correlation of the three studied FX pairs is documented by Figure \ref{corrdiff}, which confirms that co-jumps have a substantial impact on total correlations in the Asian session as the correlation difference is highest. In the US session, total correlation of the CHF--EUR pair is most of the time increased by co-jumps. The EU session, however, exhibits only a marginal role of co-jumps in the correlation structure for the GBP--CHF and GBP--EUR pairs, with the exception of the CHF--EUR pair at the end of the period.

This result is puzzling since we have previously documented almost no co-jumps during Asian session. Hence, the result could be possibly biased due to very small number of observations when co-jumps occur. To support the findings, we look at the probability that the difference between total and continuous correlations will be positive, conditional on information in co-jumps
$$Pr\left\{\widehat{corr}_T^{(t)} \ge \widehat{corr}_T^{(c)}\middle|\widehat{CJ}_T\right\}$$
being equal to 1/2 under the null of no effect of co-jumps when total and continuous correlations are equal. To connect the co-jump events with the positive difference in correlations, we look at coefficients of the following logistic regression
\begin{equation}
\label{logitcojump}
Pr\left\{\widehat{corr}_T^{(t)} \ge \widehat{corr}_T^{(c)} |\widehat{CJ}_T\right\} = \frac{1}{1+e^{-\theta}},
\end{equation}
where $\theta = \beta_0 + \beta_1 \widehat{CJ}_T$. If $\beta_0 = \beta_1 = 0$, probability will be equal to 1/2 implying that the correlations are equivalent.

Table \ref{tabcojimpactlogit} shows the results from the estimation. Estimated coefficients on the EU and U.S. sessions are jointly different from zero, while this does not hold for the Asian session. When computing the probabilities on the whole trading day, we again jointly reject the insignificance of parameters.

\begin{table}[!t]
\caption{Impact of co-jumps. Table \ref{tabcojimpactlogit} shows estimated coefficients from the regression $Pr\left\{\widehat{corr}_T \ge \widehat{corr}_T^{(c)} |\widehat{CJ}_T\right\} = 1/\left(1+e^{-(\beta_0 + \beta_1 \widehat{CJ}_T)}\right)$. All cases when $\mathcal{H}_0: \beta_0 = 0 \cap \beta_1 = 0$ is rejected using Wald test with heteroskedasticity consistent with White's covariance estimator, and where $Pr\left\{\widehat{corr}_T \ge \widehat{corr}_T^{(c)} |\widehat{CJ}_T\right\}>1/2$ are indicated by bold.}
\small
\begin{center}
\begin{tabular}{lrcccc}
\toprule
\multicolumn{2}{c}{} & Asia& EU& U.S.& Total \\
\cmidrule(r){2-6}
\multirow{4}{*}{GBP--CHF}
&$\beta_0$&  1.083 & \textbf{0.165} & \textbf{0.458} & \textbf{0.556} \\
&$\beta_1$& 37.620 & \textbf{19.900} & \textbf{13.650} & \textbf{11.520} \\
&Pseudo $R^2$& 0.006 & 0.027 & 0.012 & 0.018 \\
 \cmidrule(r){2-6}
\multirow{4}{*}{GBP--EUR}
&$\beta_0$&   1.165 & \textbf{0.135} & \textbf{0.597} & \textbf{0.653} \\
&$\beta_1$& 23.480 & \textbf{21.060} & \textbf{24.050} & \textbf{15.960} \\
&Pseudo $R^2$&   0.005 & 0.030 & 0.023 & 0.031 \\
 \cmidrule(r){2-6}
\multirow{4}{*}{ CHF--EUR}
&$\beta_0$& 1.417 & \textbf{0.886} & \textbf{1.032} & \textbf{1.290} \\
&$\beta_1$&  27.420 & \textbf{15.560} & \textbf{12.940} & \textbf{11.110} \\
&Pseudo $R^2$&   0.009 & 0.025 & 0.013 & 0.023 \\
\bottomrule
\end{tabular}
\end{center}
\label{tabcojimpactlogit}
\end{table}

We can conclude that we find significant impact of co-jumps on correlations except for the Asian session. The co-jumps, when present, are significant part of total correlations between the studied currencies. This result is consistent with earlier findings that correlation during the Asian session are generally lower in comparison to the other sessions.

\section{Conclusion}

Although most studies have focused on the precise estimation of integrated covariance structures, the role of co-jumps in overall correlations remains incompletely understood. In this paper, we investigate how co-jumps impact covariance structures in the currency markets. For this purpose, we develop a new jump wavelet covariance estimator and bootstrap testing procedure to identify co-jumps. The newly proposed methodology builds on the current co-jump literature by allowing for precise jump and co-jump detection while minimizing the identification of false co-jumps resulting from the occurrence of large idiosyncratic jumps.

While we are the first to explore usefulness of wavelet decomposition in estimating covariance and co-jumps, we view our main contribution in documenting how precisely localized co-jumps impact correlation structures in the currency market. In a real-world application, we document how co-jumps significantly influence correlations in currency markets. Next, we study the behavior of co-jumps during Asian, European and U.S. trading sessions. Our results show that the proportion of co-jumps relative to the covariance increased during 2012 -- 2015. Hence, the impact of co-jumps on correlations increased, and appropriately estimating co-jumps is becoming a crucial step in understanding dependence in currency markets. 

\section*{References}
\bibliography{thesis}

\begin{thebibliography}{}

\bibitem[\protect\citeauthoryear{A\"{i}t-Sahalia, Fan, and Xiu}{A\"{i}t-Sahalia
  et~al.}{2010}]{sahalia2011}
A\"{i}t-Sahalia, Y., J.~Fan, and D.~Xiu (2010).
\newblock High frequency covariance estimates with noisy and asynchronous
  financial data.
\newblock {\em Journal of the American Statistical Association\/}~{\em
  105\/}(492), 1504--1517.

\bibitem[\protect\citeauthoryear{A\"{i}t-Sahalia and Jacod}{A\"{i}t-Sahalia and
  Jacod}{2009}]{sahaliajacod2009}
A\"{i}t-Sahalia, Y. and J.~Jacod (2009).
\newblock Testing for jumps in a discretely observed process.
\newblock {\em The Annals of Statistics\/}~{\em 37\/}(1), 184--222.

\bibitem[\protect\citeauthoryear{A{\"\i}t-Sahalia and Jacod}{A{\"\i}t-Sahalia
  and Jacod}{2012}]{ait2012analyzing}
A{\"\i}t-Sahalia, Y. and J.~Jacod (2012).
\newblock Analyzing the spectrum of asset returns: Jump and volatility
  components in high frequency data.
\newblock {\em Journal of Economic Literature\/}~{\em 50\/}(4), 1007--1050.

\bibitem[\protect\citeauthoryear{Andersen, Bollerslev, Diebold, and
  Labys}{Andersen et~al.}{2003}]{abdl2003}
Andersen, T., T.~Bollerslev, F.~Diebold, and P.~Labys (2003).
\newblock Modeling and forecasting realized volatility.
\newblock {\em Econometrica\/}~{\em 71\/}(2), 579--625.

\bibitem[\protect\citeauthoryear{Andersen, Bollerslev, and Diebold}{Andersen
  et~al.}{2007}]{abd2007}
Andersen, T.~G., T.~Bollerslev, and F.~X. Diebold (2007).
\newblock Roughing it up: Including jump components in the measurement,
  modeling, and forecasting of return volatility.
\newblock {\em Review of Economics and Statistics\/}~{\em 89\/}(4), 701--720.

\bibitem[\protect\citeauthoryear{Andersen, Bollerslev, Frederiksen, and
  {\O}rregaard~Nielsen}{Andersen et~al.}{2010}]{andersen2010continuous}
Andersen, T.~G., T.~Bollerslev, P.~Frederiksen, and M.~{\O}rregaard~Nielsen
  (2010).
\newblock Continuous-time models, realized volatilities, and testable
  distributional implications for daily stock returns.
\newblock {\em Journal of Applied Econometrics\/}~{\em 25\/}(2), 233--261.

\bibitem[\protect\citeauthoryear{Antoniou and Gustafson}{Antoniou and
  Gustafson}{1999}]{antoniou1999}
Antoniou, I. and K.~Gustafson (1999).
\newblock Wavelets and stochastic processes.
\newblock {\em Mathematics and Computers in Simulation\/}~{\em 49\/}(1-2),
  81--104.

\bibitem[\protect\citeauthoryear{Bandi, Perron, Tamoni, and Tebaldi}{Bandi
  et~al.}{2016}]{bandi2016economic}
Bandi, F., B.~Perron, A.~Tamoni, and C.~Tebaldi (2016).
\newblock Economic uncertainty and predictability.
\newblock {\em Available at SSRN\/}.

\bibitem[\protect\citeauthoryear{Bandi and Tamoni}{Bandi and
  Tamoni}{2017}]{bandi2015business}
Bandi, F.~M. and A.~Tamoni (2017).
\newblock The horizon of systematic risk: a new beta representation.
\newblock {\em Available at SSRN\/}.

\bibitem[\protect\citeauthoryear{Barndorff-Nielsen, Hansen, Lunde, and
  Shephard}{Barndorff-Nielsen et~al.}{2011}]{BN2011}
Barndorff-Nielsen, O., P.~Hansen, A.~Lunde, and N.~Shephard (2011).
\newblock Multivariate realised kernels: Consistent positive semi-definite
  estimators of the covariation of equity prices with noise and non-synchronous
  trading.
\newblock {\em Journal of Econometrics\/}~{\em 162\/}(2), 149--169.

\bibitem[\protect\citeauthoryear{Barndorff-Nielsen and
  Shephard}{Barndorff-Nielsen and Shephard}{2004a}]{barndorff2004b}
Barndorff-Nielsen, O. and N.~Shephard (2004a).
\newblock Econometric analysis of realized covariation: High frequency based
  covariance, regression, and correlation in financial economics.
\newblock {\em Econometrica\/}~{\em 72\/}(3), 885--925.

\bibitem[\protect\citeauthoryear{Barndorff-Nielsen and
  Shephard}{Barndorff-Nielsen and Shephard}{2004b}]{barndorff2004}
Barndorff-Nielsen, O. and N.~Shephard (2004b).
\newblock Power and bipower variation with stochastic volatility and jumps.
\newblock {\em Journal of Financial Econometrics\/}~{\em 2\/}(1), 1--48.

\bibitem[\protect\citeauthoryear{Barndorff-Nielsen and
  Shephard}{Barndorff-Nielsen and Shephard}{2006}]{barndorff2006}
Barndorff-Nielsen, O. and N.~Shephard (2006).
\newblock Econometrics of testing for jumps in financial economics using
  bipower variation.
\newblock {\em Journal of Financial Econometrics\/}~{\em 4\/}(1), 1--30.

\bibitem[\protect\citeauthoryear{Barunik, Krehlik, and Vacha}{Barunik
  et~al.}{2016}]{barunikjwg}
Barunik, J., T.~Krehlik, and L.~Vacha (2016).
\newblock Modeling and forecasting exchange rate volatility in time-frequency
  domain.
\newblock {\em European Journal of Operational Reseach\/}~{\em 251\/}(1),
  329--340.

\bibitem[\protect\citeauthoryear{Barunik and Vacha}{Barunik and
  Vacha}{2015}]{barunik}
Barunik, J. and L.~Vacha (2015).
\newblock Realized wavelet-based estimation of integrated variance and jumps in
  the presence of noise.
\newblock {\em Quantitative Finance\/}~{\em 15\/}(8), 1347--1364.

\bibitem[\protect\citeauthoryear{Bibinger and Winkelmann}{Bibinger and
  Winkelmann}{2015}]{bibinger2015econometrics}
Bibinger, M. and L.~Winkelmann (2015).
\newblock Econometrics of co-jumps in high-frequency data with noise.
\newblock {\em Journal of Econometrics\/}~{\em 184\/}(2), 361--378.

\bibitem[\protect\citeauthoryear{Bidder and Dew-Becker}{Bidder and
  Dew-Becker}{2016}]{bidder2016long}
Bidder, R. and I.~Dew-Becker (2016).
\newblock Long-run risk is the worst-case scenario.
\newblock {\em The American Economic Review\/}~{\em 106\/}(9), 2494--2527.

\bibitem[\protect\citeauthoryear{Bollerslev, Law, and Tauchen}{Bollerslev
  et~al.}{2008}]{bollerslev2008risk}
Bollerslev, T., T.~H. Law, and G.~Tauchen (2008).
\newblock Risk, jumps, and diversification.
\newblock {\em Journal of Econometrics\/}~{\em 144\/}(1), 234--256.

\bibitem[\protect\citeauthoryear{Bollerslev, Osterrieder, Sizova, and
  Tauchen}{Bollerslev et~al.}{2013}]{bollerslev2013risk}
Bollerslev, T., D.~Osterrieder, N.~Sizova, and G.~Tauchen (2013).
\newblock Risk and return: Long-run relations, fractional cointegration, and
  return predictability.
\newblock {\em Journal of Financial Economics\/}~{\em 108\/}(2), 409--424.

\bibitem[\protect\citeauthoryear{Boons and Tamoni}{Boons and
  Tamoni}{2015}]{boons2015horizon}
Boons, M. and A.~Tamoni (2015).
\newblock Horizon-specific macroeconomic risks and the cross-section of
  expected returns.
\newblock {\em Available at SSRN\/}.

\bibitem[\protect\citeauthoryear{Boudt and Zhang}{Boudt and
  Zhang}{2015}]{boudt2015jump}
Boudt, K. and J.~Zhang (2015).
\newblock Jump robust two time scale covariance estimation and realized
  volatility budgets.
\newblock {\em Quantitative Finance\/}~{\em 15\/}(6), 1041--1054.

\bibitem[\protect\citeauthoryear{Caporin, Kolokolov, and Ren{\`o}}{Caporin
  et~al.}{2016}]{caporin2014multi}
Caporin, M., A.~Kolokolov, and R.~Ren{\`o} (2016).
\newblock Systemic co-jumps.
\newblock {\em Journal of Financial Economics (forthcoming)\/}.

\bibitem[\protect\citeauthoryear{Crouzet, Dew-Becker, and Nathanson}{Crouzet
  et~al.}{2017}]{crouzet2017model}
Crouzet, N., I.~Dew-Becker, and C.~G. Nathanson (2017).
\newblock A model of multi-frequency trade.
\newblock {\em Available at SSRN\/}.

\bibitem[\protect\citeauthoryear{Daubechies}{Daubechies}{1992}]{Daubechies1992}
Daubechies, I. (1992).
\newblock {\em Ten lectures on wavelets}.
\newblock SIAM.

\bibitem[\protect\citeauthoryear{Dew-Becker}{Dew-Becker}{2017}]{dew2017risky}
Dew-Becker, I. (2017).
\newblock How risky is consumption in the long-run? benchmark estimates from a
  robust estimator.
\newblock {\em Review of Financial Studies\/}~{\em 30\/}(2), 631--666.

\bibitem[\protect\citeauthoryear{Dew-Becker and Giglio}{Dew-Becker and
  Giglio}{2016}]{dew2016asset}
Dew-Becker, I. and S.~Giglio (2016).
\newblock Asset pricing in the frequency domain: theory and empirics.
\newblock {\em Review of Financial Studies\/}~{\em 29\/}(8), 2029--2068.

\bibitem[\protect\citeauthoryear{Donoho and Johnstone}{Donoho and
  Johnstone}{1994}]{donoho}
Donoho, D.~L. and I.~M. Johnstone (1994).
\newblock Ideal spatial adaptation by wavelet shrinkage.
\newblock {\em Biometrica\/}~{\em 81\/}(3), 425--455.

\bibitem[\protect\citeauthoryear{Dovonon, Gon{\c{c}}alves, Hounyo, and
  Meddahi}{Dovonon et~al.}{2014}]{dovonon2014bootstrapping}
Dovonon, P., S.~Gon{\c{c}}alves, U.~Hounyo, and N.~Meddahi (2014).
\newblock Bootstrapping high-frequency jump tests.
\newblock Discussion paper, Toulouse School of Economics.

\bibitem[\protect\citeauthoryear{Epps}{Epps}{1979}]{epps}
Epps, T.~W. (1979).
\newblock Comovements in stock prices in the very short run.
\newblock {\em Journal of the American Statistical Association\/}~{\em
  74\/}(366a), 291--298.

\bibitem[\protect\citeauthoryear{Fan and Wang}{Fan and
  Wang}{2007}]{fanwang2008}
Fan, J. and Y.~Wang (2007).
\newblock Multi-scale jump and volatility analysis for high-frequency financial
  data.
\newblock {\em Journal of the American Statistical Association\/}~{\em
  102\/}(480), 1349--1362.

\bibitem[\protect\citeauthoryear{Gen{\c c}ay, Sel{\c c}uk, and Whitcher}{Gen{\c
  c}ay et~al.}{2002}]{Gencay2002}
Gen{\c c}ay, R., F.~Sel{\c c}uk, and B.~Whitcher (2002).
\newblock {\em An Introduction to Wavelets and Other Filtering Methods in
  Finance and Economics.}
\newblock Academic Press.

\bibitem[\protect\citeauthoryear{Gen{\c{c}}ay, Sel{\c{c}}uk, and
  Whitcher}{Gen{\c{c}}ay et~al.}{2005}]{genccay2005multiscale}
Gen{\c{c}}ay, R., F.~Sel{\c{c}}uk, and B.~Whitcher (2005).
\newblock Multiscale systematic risk.
\newblock {\em Journal of International Money and Finance\/}~{\em 24\/}(1),
  55--70.

\bibitem[\protect\citeauthoryear{Gilder, Shackleton, and Taylor}{Gilder
  et~al.}{2014}]{gilder2014cojumps}
Gilder, D., M.~B. Shackleton, and S.~J. Taylor (2014).
\newblock Cojumps in stock prices: Empirical evidence.
\newblock {\em Journal of Banking \& Finance\/}~{\em 40}, 443--459.

\bibitem[\protect\citeauthoryear{Gnabo, Hvozdyk, and Lahaye}{Gnabo
  et~al.}{2014}]{gnabo2014system}
Gnabo, J.-Y., L.~Hvozdyk, and J.~Lahaye (2014).
\newblock System-wide tail comovements: A bootstrap test for cojump
  identification on the {S}\&{P} 500, {US} bonds and currencies.
\newblock {\em Journal of International Money and Finance\/}~{\em 48\/}(A),
  147--174.

\bibitem[\protect\citeauthoryear{Griffin and Oomen}{Griffin and
  Oomen}{2011}]{griffin}
Griffin, J. and R.~Oomen (2011).
\newblock Covariance measurement in the presence of non-synchronous trading and
  market microstructure noise.
\newblock {\em Journal of Econometrics\/}~{\em 160\/}(1), 58--68.

\bibitem[\protect\citeauthoryear{Hayashi and Yoshida}{Hayashi and
  Yoshida}{2005}]{hayashiyoshida}
Hayashi, T. and N.~Yoshida (2005).
\newblock On covariance estimation of non-synchronously observed diffusion
  processes.
\newblock {\em Bernoulli\/}~{\em 11\/}(2), 359--379.

\bibitem[\protect\citeauthoryear{Jacod and Todorov}{Jacod and
  Todorov}{2009}]{jacod2009testing}
Jacod, J. and V.~Todorov (2009).
\newblock Testing for common arrivals of jumps for discretely observed
  multidimensional processes.
\newblock {\em The Annals of Statistics\/}~{\em 37\/}(4), 1792--1838.

\bibitem[\protect\citeauthoryear{Jawadi, Louhichi, and Cheffou}{Jawadi
  et~al.}{2015}]{jawadi2015testing}
Jawadi, F., W.~Louhichi, and A.~I. Cheffou (2015).
\newblock Testing and modeling jump contagion across international stock
  markets: A nonparametric intraday approach.
\newblock {\em Journal of Financial Markets\/}~{\em 26}, 64--84.

\bibitem[\protect\citeauthoryear{Lahaye, Laurent, and Neely}{Lahaye
  et~al.}{2011}]{Lahaye2011}
Lahaye, J., S.~Laurent, and C.~J. Neely (2011).
\newblock Jumps, cojumps and macro announcements.
\newblock {\em Journal of Applied Econometrics\/}~{\em 26\/}(6), 893--921.

\bibitem[\protect\citeauthoryear{Lee and Mykland}{Lee and
  Mykland}{2008}]{lee2008}
Lee, S. and P.~A. Mykland (2008).
\newblock Jumps in financial markets: a new nonpara- metric test and jump
  dynamics.
\newblock {\em The Review of Financial Studies\/}~{\em 21}, 2525--2563.

\bibitem[\protect\citeauthoryear{Li and Zhang}{Li and
  Zhang}{2017}]{li2017short}
Li, J. and H.~H. Zhang (2017).
\newblock Short-run and long-run consumption risks, dividend processes, and
  asset returns.
\newblock {\em Review of Financial Studies\/}~{\em 30\/}(2), 588--630.

\bibitem[\protect\citeauthoryear{Mallat}{Mallat}{1998}]{Mallat98}
Mallat, S. (1998).
\newblock {\em A wavelet tour of signal processing}.
\newblock Academic Press.

\bibitem[\protect\citeauthoryear{Mancini and Gobbi}{Mancini and
  Gobbi}{2012}]{mancini2012identifying}
Mancini, C. and F.~Gobbi (2012).
\newblock Identifying the brownian covariation from the co-jumps given discrete
  observations.
\newblock {\em Econometric Theory\/}~{\em 28\/}(02), 249--273.

\bibitem[\protect\citeauthoryear{Novotn{\'y}, Petrov, and Urga}{Novotn{\'y}
  et~al.}{2015}]{novotny2015trading}
Novotn{\'y}, J., D.~Petrov, and G.~Urga (2015).
\newblock Trading price jump clusters in foreign exchange markets.
\newblock {\em Journal of Financial Markets\/}~{\em 24}, 66--92.

\bibitem[\protect\citeauthoryear{Otrok, Ravikumar, and Whiteman}{Otrok
  et~al.}{2002}]{otrok2002habit}
Otrok, C., B.~Ravikumar, and C.~H. Whiteman (2002).
\newblock Habit formation: a resolution of the equity premium puzzle?
\newblock {\em Journal of Monetary Economics\/}~{\em 49\/}(6), 1261--1288.

\bibitem[\protect\citeauthoryear{Patton and Sheppard}{Patton and
  Sheppard}{2009}]{patton2009evaluating}
Patton, A.~J. and K.~Sheppard (2009).
\newblock Evaluating volatility and correlation forecasts.
\newblock In {\em Handbook of financial time series}, pp.\  801--838. Springer.

\bibitem[\protect\citeauthoryear{Percival and Mofjeld}{Percival and
  Mofjeld}{1997}]{PercivalMofjeld1997}
Percival, D.~B. and H.~Mofjeld (1997).
\newblock Analysis of subtidal coastal sea level fluctuations using wavelets.
\newblock {\em Journal of the American Statistical Association\/}~{\em
  92\/}(439), 886--880.

\bibitem[\protect\citeauthoryear{Percival and Walden}{Percival and
  Walden}{2000}]{PercivalWalden2000}
Percival, D.~B. and A.~T. Walden (2000).
\newblock {\em Wavelet Methods for Time series Analysis}.
\newblock Cambridge University Press.

\bibitem[\protect\citeauthoryear{Protter}{Protter}{1992}]{protter}
Protter, P. (1992).
\newblock {\em Stochastic integration and differential equations: A new
  approach}.
\newblock New York: Springer-Verlag.

\bibitem[\protect\citeauthoryear{Serroukh and Walden}{Serroukh and
  Walden}{2000a}]{serroukhwalden2000a}
Serroukh, A. and A.~Walden (2000a).
\newblock Wavelet scale analysis of bivariate time series i: motivation and
  estimation.
\newblock {\em Journal of Nonparametric Statistics\/}~{\em 13\/}(1), 1--36.

\bibitem[\protect\citeauthoryear{Serroukh and Walden}{Serroukh and
  Walden}{2000b}]{serroukhwalden2000b}
Serroukh, A. and A.~Walden (2000b).
\newblock Wavelet scale analysis of bivariate time series ii: statistical
  properties for linear processes.
\newblock {\em Journal of Nonparametric Statistics\/}~{\em 13\/}(1), 37--56.

\bibitem[\protect\citeauthoryear{Varneskov}{Varneskov}{2016}]{varneskov2015flat}
Varneskov, R.~T. (2016).
\newblock Flat-top realized kernel estimation of quadratic covariation with
  non-synchronous and noisy asset prices.
\newblock {\em Journal of Business \& Economic Statistics\/}~{\em 34\/}(1),
  1--22.

\bibitem[\protect\citeauthoryear{Voev and Lunde}{Voev and
  Lunde}{2007}]{voevlunde}
Voev, V. and A.~Lunde (2007).
\newblock Integrated covariance estimation using high-frequency data in the
  presence of noise.
\newblock {\em Journal of Financial Econometrics\/}~{\em 5\/}(1), 68--104.

\bibitem[\protect\citeauthoryear{Wang}{Wang}{1995}]{wang95}
Wang, Y. (1995).
\newblock Jump and sharp cusp detection via wavelets.
\newblock {\em Biometrika\/}~{\em 82\/}(2), 385--397.

\bibitem[\protect\citeauthoryear{Whitcher, Guttorp, and Percival}{Whitcher
  et~al.}{1999}]{WGPTech1999}
Whitcher, B., P.~Guttorp, and D.~B. Percival (1999).
\newblock Mathematical background for wavelets estimators for cross covariance
  and cross correlation.
\newblock Technical Report~38, National Resource Centre for Supplementary
  Education.

\bibitem[\protect\citeauthoryear{Whitcher, Guttorp, and Percival}{Whitcher
  et~al.}{2000}]{WhitcherGP2000}
Whitcher, B., P.~Guttorp, and D.~B. Percival (2000).
\newblock Wavelet analysis of covariance with application to atmosferic time
  series.
\newblock {\em Journal of Geophysical Research\/}~{\em 105\/}(D11), 941--962.

\bibitem[\protect\citeauthoryear{Xue, Gen{\c{c}}ay, and Fagan}{Xue
  et~al.}{2014}]{xue2014jump}
Xue, Y., R.~Gen{\c{c}}ay, and S.~Fagan (2014).
\newblock Jump detection with wavelets for high-frequency financial time
  series.
\newblock {\em Quantitative Finance\/}~{\em 14\/}(8), 1427--1444.

\bibitem[\protect\citeauthoryear{Yu}{Yu}{2012}]{yu2012using}
Yu, J. (2012).
\newblock Using long-run consumption-return correlations to test asset pricing
  models.
\newblock {\em Review of Economic Dynamics\/}~{\em 15\/}(3), 317--335.

\bibitem[\protect\citeauthoryear{Zhang}{Zhang}{2011}]{Zhang2011}
Zhang, L. (2011).
\newblock Estimating covariation: Epps effect, microstructure noise.
\newblock {\em Journal of Econometrics\/}~{\em 160\/}(1), 33--47.

\bibitem[\protect\citeauthoryear{Zhang, Mykland, and A\"{i}t-Sahalia}{Zhang
  et~al.}{2005}]{zhang2005}
Zhang, L., P.~Mykland, and Y.~A\"{i}t-Sahalia (2005).
\newblock A tale of two time scales: Determining integrated volatility with
  noisy high frequency data.
\newblock {\em Journal of the American Statistical Association\/}~{\em
  100\/}(472), 1394--1411.

\end{thebibliography}
\bibliographystyle{chicago}

\newpage
\section{Appendix: Supplementary Tables}


\begin{table}[!h]
\caption{Descriptive Statistics for British Pound (GBP), Swiss franc (CHF), and Euro (EUR) futures logarithmic price returns. The sample period runs between January 5, 2007 and July 3, 2015. Descriptive statistics is reported for the 1-minute, and 5-minute frequency of intraday returns, respectively. Means are scaled by $10^7$, and standard deviations are scaled by $10^4$.}
\small
\begin{center}
\begin{tabular}{l ccccccc}
\toprule
& \multicolumn{3}{c}{1 minute} & & \multicolumn{3}{c}{5 minute} \\
\cmidrule(r){2-4} \cmidrule(r){6-8}
& GBP & CHF & EUR & & GBP & CHF & EUR \\
\cmidrule(r){2-4} \cmidrule(r){6-8}
Mean & 0.388  &  1.447 & 0.556 & & 2.209 & 7.123 & 2.844 \\
Minimum & -0.011 & -0.042 & -0.014 & & -0.012 & -0.055 & -0.013 \\
Maximum & 0.010 & 0.096 & 0.017 & & 0.014 & 0.095 & 0.014 \\
Std. Dev. & 1.899 & 2.167 & 1.903 & & 4.059 & 4.635 & 4.053 \\
Skewness & -0.051 & -3.479 & 0.381 &  & -0.060 & -5.239 & 0.103 \\
Kurtosis & 55.512 & 605.088 & 86.283 & &  27.402 & 542.866 & 27.381 \\
\bottomrule
\end{tabular}
\end{center}
\label{tabdesc}
\end{table}

\begin{table}[!h]
\caption{Continuous covariation bias (variance in parenthesis) $\times 10^4$ of all estimators from 10,000 simulations of the jump-diffusion model with $\epsilon_1=0$, $\epsilon_2=0.0015$, zero and one co-jump (CJ), and zero and one independent jump (IJ). RC -- realized covariance, BC -- bipower covariance, TSCV -- two-scale realized covariance, MRK -- multivariate realized, JWC -- jump wavelet covariance with sampling at 1-min, 5-min, 30-min and 1-hour intervals.}
\scriptsize
\centering
\ra{1}
\begin{tabular}{lrrrrrrr}
\toprule
& & & \multicolumn{1}{c}{\textbf{RC}} &\multicolumn{1}{c}{\textbf{BC}} &\multicolumn{1}{c}{\textbf{TSCV}} &\multicolumn{1}{c}{\textbf{MRK}} & \multicolumn{1}{c}{\textbf{JWC}} \\
\midrule
& & & \multicolumn{5}{c}{Z e r o  N o i s e ($\epsilon_1$)} \\
\midrule
\multirow{8}{*}{\rotatebox{90}{\mbox{Zero IJ}}} & \multirow{4}{*}{Zero CJ}& 1-min &-0.001 (0.015) & -0.002 (0.017) & -0.005 (0.013) & -0.006 (0.042) & -0.005 (0.013) \\
& & 5-min &0.001 (0.035) & -0.002 (0.040) & -0.002 (0.029) & -0.008 (0.069) & -0.002 (0.029) \\
& & 30-min &-0.001 (0.085) & -0.015 (0.090) & -0.015 (0.067) & -0.040 (0.112) & -0.015 (0.067) \\
& & 1-hour &0.002 (0.124) & -0.032 (0.124) & -0.030 (0.091) & -0.080 (0.129) & -0.030 (0.091) \\
\cmidrule{3-8}
 & \multirow{4}{*}{One CJ}& 1-min &0.990 (1.786) & 0.047 (0.089) & 0.969 (1.755) & 0.982 (1.805) & -0.004 (0.012) \\
& & 5-min &0.988 (1.811) & 0.107 (0.245) & 0.962 (1.772) & 0.960 (1.834) & -0.005 (0.029) \\
& & 30-min &1.019 (2.041) & 0.241 (0.577) & 0.895 (1.705) & 0.743 (1.617) & -0.018 (0.065) \\
& & 1-hour &1.001 (1.925) & 0.272 (0.745) & 0.753 (1.564) & 0.444 (1.335) & -0.036 (0.090) \\
\cmidrule{3-8}
\multirow{8}{*}{\rotatebox{90}{\mbox{One IJ}}} & \multirow{4}{*}{Zero CJ}& 1-min &-0.003 (0.042) & 0.035 (0.042) & -0.006 (0.036) & -0.000 (0.155) & -0.004 (0.012) \\
& & 5-min &-0.006 (0.115) & 0.063 (0.093) & -0.008 (0.090) & -0.014 (0.218) & -0.005 (0.028) \\
& & 30-min &0.012 (0.326) & 0.097 (0.209) & -0.007 (0.266) & -0.021 (0.467) & -0.014 (0.066) \\
& & 1-hour &-0.008 (0.568) & 0.069 (0.341) & -0.038 (0.384) & -0.096 (0.547) & -0.035 (0.090) \\
\cmidrule{3-8}
 & \multirow{4}{*}{One CJ}& 1-min &0.926 (1.624) & 0.084 (0.107) & 0.907 (1.593) & 0.917 (1.632) & -0.005 (0.012) \\
& & 5-min &1.002 (1.795) & 0.197 (0.343) & 0.988 (1.781) & 0.968 (1.860) & -0.005 (0.028) \\
& & 30-min &1.012 (1.892) & 0.417 (0.758) & 0.910 (1.768) & 0.771 (1.800) & -0.018 (0.069) \\
& & 1-hour &1.013 (2.097) & 0.493 (1.113) & 0.797 (1.730) & 0.469 (1.586) & -0.038 (0.091) \\
\midrule
& & & \multicolumn{5}{c}{N o i s e ($\epsilon_2$)} \\
\midrule
\multirow{8}{*}{\rotatebox{90}{\mbox{Zero IJ}}} & \multirow{4}{*}{Zero CJ}& 1-min &0.000 (0.015) & -0.000 (0.017) & -0.004 (0.013) & -0.002 (0.045) & -0.004 (0.013) \\
& & 5-min &-0.002 (0.035) & -0.004 (0.040) & -0.005 (0.028) & -0.009 (0.069) & -0.005 (0.028) \\
& & 30-min &0.004 (0.091) & -0.016 (0.095) & -0.015 (0.071) & -0.036 (0.130) & -0.015 (0.071) \\
& & 1-hour &-0.000 (0.124) & -0.036 (0.125) & -0.036 (0.087) & -0.086 (0.123) & -0.036 (0.087) \\
\cmidrule{3-8}
 & \multirow{4}{*}{One CJ}& 1-min &1.016 (1.745) & 0.047 (0.068) & 0.993 (1.710) & 0.999 (1.739) & -0.005 (0.013) \\
& & 5-min &0.882 (1.597) & 0.099 (0.252) & 0.866 (1.605) & 0.874 (1.691) & -0.004 (0.028) \\
& & 30-min &1.024 (1.850) & 0.261 (0.632) & 0.948 (1.774) & 0.831 (1.838) & -0.018 (0.062) \\
& & 1-hour &0.982 (1.834) & 0.292 (0.719) & 0.789 (1.615) & 0.490 (1.371) & -0.035 (0.093) \\
\cmidrule{3-8}
\multirow{8}{*}{\rotatebox{90}{\mbox{One IJ}}} & \multirow{4}{*}{Zero CJ}& 1-min &0.001 (0.049) & 0.037 (0.045) & -0.003 (0.039) & -0.001 (0.196) & -0.004 (0.012) \\
& & 5-min &0.007 (0.094) & 0.068 (0.099) & -0.001 (0.084) & -0.014 (0.248) & -0.005 (0.029) \\
& & 30-min &0.015 (0.362) & 0.097 (0.211) & 0.002 (0.268) & -0.030 (0.523) & -0.018 (0.066) \\
& & 1-hour &0.017 (0.536) & 0.072 (0.307) & -0.028 (0.370) & -0.074 (0.449) & -0.033 (0.092) \\
\cmidrule{3-8}
 & \multirow{4}{*}{One CJ}& 1-min &0.832 (1.443) & 0.076 (0.084) & 0.815 (1.418) & 0.831 (1.472) & -0.005 (0.012) \\
& & 5-min &1.042 (1.818) & 0.228 (0.473) & 1.031 (1.806) & 1.015 (1.995) & -0.004 (0.029) \\
& & 30-min &0.977 (1.865) & 0.448 (0.782) & 0.886 (1.678) & 0.763 (1.704) & -0.018 (0.067) \\
& & 1-hour &0.993 (1.957) & 0.501 (1.080) & 0.812 (1.698) & 0.515 (1.617) & -0.037 (0.088) \\
\bottomrule
\label{simul_cov1}
\end{tabular}
\end{table}

\clearpage
\pagebreak
\section{Mathematical Appendix}
\appendix

\section{Decomposition of quadratic covariation}
Using the continuous wavelet transform we can decompose price (return) process $(\mathbf{Y}_t)_{t\in[0,T]}$ (defined in Section \ref{est}) into various frequency scales. Let us start with wavelet decomposition of the quadratic variation on the diagonal terms in the covariance matrix $\bm{QV}$. The quadratic variation over a fixed time interval $[0\le t\le T]$ associated with $\mathbf{Y}_t=\left(Y_{t,\ell_1},\ldots,Y_{t,\ell_d}\right)'\in L^2(\mathbb{R})$ can be written as
\begin{equation}
\label{wvar}
QV_{\ell,\ell} = \frac{1}{C_\psi}\int_{0}^{\infty} \left[\int_{-\infty}^{\infty} \vert W_{j,k}^{\ell}\vert^2 \mathrm{d}k \right]\frac{\mathrm{d}j}{j^2},
\end{equation}
where $W_{j,k}^{\ell}$ is the continuous wavelet transform with respect to a wavelet $\psi_{j,k}(t)=\vert j \vert ^{-1/2}\psi \left( \frac{t-k}{j}\right)$ defined as:
\begin{equation}
W_{j,k}^{\ell}=\vert j \vert ^{-1/2}\int_{0}^{T}\overline{\psi\left(\frac{t-k}{j}\right)} \Delta Y_{t,\ell} \mathrm{d}t,
\end{equation}
where $\Delta Y_{t,\ell} = (\Delta_1 Y_{t,\ell},\ldots,\Delta_N
Y_{t,\ell})$ are intraday returns, $k$ denotes a specific time position in a day, whereas $j$ is a scale (related to frequency) of wavelet $\psi$, and the bar denotes complex
conjugation\footnote{For more details about the continuous wavelet transform, see \cite{Daubechies1992}.}. Eq.(\ref{wvar}) shows how the
quadratic variation of a process $Y_{t,\ell}$ can be decomposed by the
wavelet transform. Furthermore, we can generalize this result to a quadratic covariation. If $(Y_{t,\ell_1},Y_{t,\ell_2})$ belong to $L^2(\mathbb{R})$ and have a continuous wavelet transform, then the quadratic covariation can be decomposed by wavelets in a similar manner as
\begin{equation}
\label{cwcov}
QV_{\ell_1,\ell_2}=\frac{1}{C_\psi}\int_{0}^{\infty} \left[\int_{-\infty}^{\infty} W_{j,k}^{\ell_1} \overline{W_{j,k}^{\ell_2}} \mathrm{d}k \right]\frac{\mathrm{d}j}{j^2}.
\end{equation}
Eq.(\ref{cwcov}) is a starting point for the construction of a wavelet estimator of quadratic covariation. The term $\int_{-\infty}^{\infty} W_{j,k}^{\ell_1} \overline{W_{j,k}^{\ell_2}} \mathrm{d}k$ expresses the quadratic covariation at a particular scale $j$, whereas the other integral sums all of the available scales $j$. Using this representation, we can know the exact contribution of each scale to the overall quadratic covariation measure.

\section{Discrete wavelet transform}
\label{dwt}

Here, we briefly introduce a discrete version of the wavelet transform. We use a special form of the discrete wavelet transform called the maximal overlap discrete wavelet transform (MODWT). We demonstrate the application of the discrete-type wavelet transform on a stochastic process using the pyramid algorithm \citep{Mallat98}. This method is based on filtering time series (or stochastic process) with MODWT wavelet filters and then filtering the output again to obtain other wavelet scales. Using the MODWT procedure, we obtain wavelet and scaling coefficients that decompose analyzed stochastic processes into frequency bands. Ror more details about discrete wavelet transforms and their applications, see \cite{PercivalMofjeld1997}, \cite{PercivalWalden2000}, and \cite{Gencay2002}.

The pyramid algorithm has several stages, and the number of stages
depends on the maximal level of decomposition $\mathcal{J}^m$. Let us
begin with the first stage. The wavelet coefficients at the first
scale $(j=1)$ are obtained via the circular filtering of time series $Y_{t,\ell}$ using the MODWT wavelet and scaling filters $h_{1,l}$ and $g_{1,l}$ \citep{PercivalWalden2000} :
\begin{equation}
\label{GrindEQ__13_}
\mathcal{W}_{1,t}^{\ell}\equiv \sum^{L{\rm -}{\rm 1}}_{l{\rm =0}}{}h_{1,l}Y_{{\rm (}t{\rm -}l\ modN{\rm )},\ell} \ \ \ \ \mathcal{V}_{1,t}^{\ell}\equiv \sum^{L{\rm -}{\rm 1}}_{l{\rm =0}}{}g_{1,l}Y_{{\rm (}t{\rm -}l\ modN{\rm )},\ell}.
\end{equation}
In the second step, the algorithm uses the scaling coefficients $\mathcal{V}_{1,t}^{\ell}$ instead of $Y_{t,\ell}$. The wavelet and scaling filters have a width $L_j=2^{j-1}\left(L-1\right)+1$. After filtering, we obtain the wavelet coefficients at scale $j=2$:
\begin{equation}
\label{GrindEQ__14_}
\mathcal{W}_{2,t}^{\ell}\equiv \sum^{L{\rm -}{\rm 1}}_{l{\rm =0}}{}h_{2,l}\mathcal{V}_{{\rm (1,}t{\rm -}l\ modN{\rm )}}^{\ell} \ \ \ \ \mathcal{V}_{2,t}^{\ell}\equiv \sum^{L{\rm -}{\rm 1}}_{l{\rm =0}}{}g_{2,l}\mathcal{V}_{{\rm (1,}t{\rm -}l\ modN{\rm )}}^{\ell}.
\end{equation}
The two steps of the algorithm create two vectors of the MODWT wavelet
coefficients at scales $j=1$ and $j=2$;
$\mathcal{W}_{1,t}^{\ell},\mathcal{W}_{2,t}^{\ell}$ and a vector of
the MODWT wavelet scaling coefficients at scale two
$\mathcal{V}_{2,t}^{\ell}$ that is subsequently used for further
decomposition. The vector $\mathcal{W}_{1,t}^{\ell}$ represents the
wavelet coefficients that reflect the activity at the frequency bands $f[1/4,1/2]$, $\mathcal{W}_{2,t}^{\ell}$: $f [1/8,1/4]$ and $\mathcal{V}_{2,t}^{\ell}$: $f[0,1/8]$.

The transfer function of the wavelet filter $h_l:l=0,1,\dots ,L-1$, where \textit{L} is the width of the filter, is denoted as $H(.)$. The pyramid algorithm exploits the fact that if we increase the width of the filter to $2^{j-1}\left(L-1\right)+1$, the filter with the impulse response sequence has the form:
\begin{equation}
\{h_0,\underbrace{0,\dots ,0}_{\mbox{$\scriptscriptstyle{2^{j-1}-1}$ \scriptsize{zeros}}},h_1,\underbrace{0,\dots ,0}_{\mbox{$\scriptscriptstyle{2^{j-1}-1}$ \scriptsize{zeros}}},h_{L-2},\underbrace{0,\dots ,0}_{\mbox{$\scriptscriptstyle{2^{j-1}-1}$ \scriptsize{zeros}}},h_L\},
\end{equation}
and a transfer function defined as $H\left(2^{j-1}f\right)$. Then, the
pyramid algorithm takes on the following form:
\begin{equation}
\mathcal{W}_{j,t}^{\ell}\equiv \sum^{L{\rm -}{\rm 1}}_{l{\rm =0}}{}h_l\mathcal{V}^{\ell}_{\left(j{\rm -}{\rm 1,}t-2^{j-1}l\ modN\right)}{\rm \ \ \ \ }t{\rm =0,1,\dots ,N-1,\ \ \ }
\end{equation}
\begin{equation}
\mathcal{V}_{j,t}^{\ell}\equiv \sum^{L{\rm -}{\rm 1}}_{l{\rm =0}}{}g_l\mathcal{V}^{\ell}_{\left(j{\rm -}{\rm 1,}t-2^{j-1}l\ modN\right)}{\rm \ \ \ \ }t{\rm =0,1,\dots ,N-1,\ }
\end{equation}
where in the first stage, we set $Y_t=\mathcal{V}_{0,t}^{\ell}$. After
applying the MODWT, we obtain $j\le \mathcal{J}^m\le log_{2}(N)$ vectors of wavelet coefficients and one vector of
scaling coefficients. The $j$-th level wavelet coefficients in vector
$\mathcal{W}_{j,t}^{\ell}$ represent the frequency bands $f [1/2^{j+1}{\rm ,1/}{{\rm 2}}^j{\rm ]}$, whereas the $j$-th level scaling
coefficients in vector $\mathcal{V}_{j,t}^{\ell}$ represent $f[0,1/2^{j+1}]$. In our analysis, we use the MODWT with the Daubechies
wavelet filter D(4) and reflecting boundary conditions.

\section{Wavelet covariance}
\label{awc}

In this Section we define the wavelet covariance which is a crucial concept for the wavelet covariance estimators. Let $(Y_{t_,\ell_1},Y_{t_,\ell_2})$ be a covariance stationary process with the square integrable spectral density functions $S^{\ell_1}(.)$, $S^{\ell_2}(.)$ and cross spectra $S^{\ell_1,\ell_2} (.)$. The wavelet covariance of $(Y_{t_,\ell_1},Y_{t_,\ell_2})$ at level $j$ is defined as:

\begin{equation}
\gamma_j^{\ell_1,\ell_2}=Cov\left( \mathcal{W}_{j,t}^{\ell_1},\, \mathcal{W}_{j,t}^{\ell_2}\right),
\end{equation}
where $\mathcal{W}_{j,t}^{\ell_1},\mathcal{W}_{j,t}^{\ell_2}$ are vectors of MODWT coefficients for $Y_{t_,\ell_1}$ and $Y_{t_,\ell_2}$, respectively. For a particular level of decomposition $\mathcal{J}^m\le log_2(T)$, the covariance of $(Y_{t_,\ell_1},Y_{t_,\ell_2})$ is a sum of the covariances of the MODWT wavelet coefficients $\gamma_j^{\ell_1,\ell_2}$ at all scales $j=1,2,\ldots,\mathcal{J}^m$ and the covariance of the scaling coefficients $\mathcal{V}_{\mathcal{J}^m,t}^{\ell}$ at scale $\mathcal{J}^m$:

\begin{equation}
\label{WaveCov}
Cov\left(Y_{t_,\ell_1},Y_{t_,\ell_2}\right)=Cov\left(\mathcal{V}_{\mathcal{J}^m,t}^{\ell_1},\, \mathcal{V}_{\mathcal{J}^m,t}^{\ell_2}\right)+\sum_{j=1}^{\mathcal{J}^m} \gamma_j^{\ell_1,\ell_2}.
\end{equation}

For process $(Y_{t_,\ell_1},Y_{t_,\ell_2})$ defined above, the estimator of a wavelet covariance at level $j$ is defined as
\begin{equation}
\widehat{\gamma}_j^{\ell_1,\ell_2}=\frac{1}{M_j}\sum_{t=L_j-1}^{N-1} \mathcal{W}_{j,t}^{\ell_1} \mathcal{W}_{j,t}^{\ell_2},
\end{equation}
where $M_j=N-L_j+1>0$ is number of the $j$-th level MODWT coefficients for both processes that are unaffected by the boundary conditions. \cite{WGPTech1999} prove that for the Gaussian process $(Y_{t_,\ell_1},Y_{t_,\ell_2})$, the MODWT  estimator of wavelet covariance is unbiased and asymptotically normally distributed.

\begin{prop}
\label{WaveCov2}
When $\mathcal{J}^m\rightarrow \infty$, the covariance of the scaling coefficients $\left(\mathcal{V}_{\mathcal{J}^m,t}^{\ell_1},\,\mathcal{V}_{\mathcal{J}^m,t}^{\ell_2}\right)$ goes to zero \citep{WGPTech1999}, and thus, we can rewrite (\ref{WaveCov}) as:
\begin{equation}
\label{WaveCov2eq}
Cov\left(Y_{t_,\ell_1},Y_{t_,\ell_2}\right)=\sum_{j=1}^\infty \gamma_j^{\ell_1,\ell_2}.
\end{equation}
\end{prop}

\textbf{Proof}:
\label{proofWaveCov}
To prove Proposition \ref{WaveCov2}, we write the covariance of the MODWT wavelet coefficients in the form:
\begin{equation}
\gamma_j^{\ell_1,\ell_2}=\int_{-1/2}^{1/2} \mathcal{H}_j (f) S^{\ell_1,\ell_2}(f)df,
\end{equation}
where $\mathcal{H}_j (f)$ denotes the squared gain function of the wavelet MODWT filter $h_j$. The covariance of the scaling coefficients at level $\mathcal{J}^m$ (the last level of decomposition):
\begin{equation}
Cov\left(\mathcal{V}^{\ell_1}_{\mathcal{J}^m,t},\, \mathcal{V}^{\ell_2}_{\mathcal{J}^m,t}\right)=\int_{-1/2}^{1/2} \mathcal{G}_J (f) S^{\ell_1,\ell_2}(f)df,
\end{equation}
where $\mathcal{G}_{\mathcal{J}^m} (f)$ denotes the squared gain function of the scaling MODWT filter $g_{\mathcal{J}^m}$, such that $\mathcal{G}_{\mathcal{J}^m} (f)\equiv \prod_{l=0}^{\mathcal{J}^m-1}\mathcal{G}(2^lf)$. When $\mathcal{H}(f)+\mathcal{G}(f)=1$ \citep{PercivalWalden2000}, the covariance decomposed by wavelets at the first level ($\mathcal{J}^m=1$) only is obtained as the sum of the wavelet and scaling MODWT coefficients' covariances,

\begin{equation}
Cov(Y_{t_,\ell_1},Y_{t_,\ell_2})=\int_{-1/2}^{1/2} \left(\mathcal{H}(f)+\mathcal{G}(f)\right) S^{\ell_1,\ell_2}(f)df=Cov\left(\mathcal{V}^{\ell_1}_{1,t},\, \mathcal{V}^{\ell_2}_{1,t}\right)+\gamma_1^{\ell_1,\ell_2}.
\end{equation}
Further, we assume that this also holds for level $\mathcal{J}^m-1$:
\begin{equation}
\label{Wconinf}
Cov(Y_{t_,\ell_1},Y_{t_,\ell_2})=Cov\left(\mathcal{V}^{\ell_1}_{\mathcal{J}^m-1,t},\, \mathcal{V}^{\ell_2}_{\mathcal{J}^m-1,t} \right)+\sum_{j=1}^{\mathcal{J}^m-1} \gamma_j^{\ell_1,\ell_2}.
\end{equation}
Following \cite{WGPTech1999}, we have
\begin{eqnarray}
\label{CovScale}
\nonumber Cov\left(\mathcal{V}^{\ell_1}_{\mathcal{J}^m-1,t},\, \mathcal{V}^{\ell_2}_{\mathcal{J}^m-1,t} \right) &=&\int_{-1/2}^{1/2} \mathcal{G}_{\mathcal{J}^m-1} (f) S^{\ell_1,\ell_2}(f)df \\
\nonumber &=&\int_{-1/2}^{1/2}\left[\prod_{l=0}^{\mathcal{J}^m-2} \mathcal{G}(2^lf)  \right] S^{\ell_1,\ell_2}(f)df\\
\nonumber &=&\int_{-1/2}^{1/2}  \left[  \mathcal{G}(2^{\mathcal{J}^m-1}f)+\mathcal{H}(2^{\mathcal{J}^m-1}f) \right]  \left[\prod_{l=0}^{\mathcal{J}^m-2} \mathcal{G}(2^lf)  \right] S^{\ell_1,\ell_2}(f)df\\
\nonumber &=&\int_{-1/2}^{1/2}  \left[  \mathcal{G}_{\mathcal{J}^m}(f)+\mathcal{H}_{\mathcal{J}^m}(f) \right]  S^{\ell_1,\ell_2}(f)df\\
&=& Cov\left(\mathcal{V}^{\ell_1}_{\mathcal{J}^m,t},\, \mathcal{V}^{\ell_2}_{\mathcal{J}^m,t}\right)+\gamma_j^{\ell_1,\ell_2},
\end{eqnarray}
which proves, by induction, the wavelet covariance decomposition of $(Y_{t_,\ell_1},Y_{t_,\ell_2})$ for a finite number of scales $\mathcal{J}^m$. We also prove that as $\mathcal{J}^m\rightarrow \infty$, the
covariance between between the scaling coefficients goes to zero; therefore, the covariance of $(Y_{t_,\ell_1},Y_{t_,\ell_2})$ depends only on the covariance of the wavelet coefficients $\gamma_j^{\ell_1,\ell_2}$. Using the result (\ref{CovScale}), we can write:
\begin{eqnarray}
Cov(\mathcal{V}^{\ell_1}_{\mathcal{J}^m-1,t},\, \mathcal{V}^{\ell_2}_{\mathcal{J}^m-1,t})&=&Cov(\mathcal{V}^{\ell_1}_{\mathcal{J}^m,t},\, \mathcal{V}^{\ell_2}_{\mathcal{J}^m,t})+\gamma_j^{\ell_1,\ell_2}\\
Cov(\mathcal{V}^{\ell_1}_{\mathcal{J}^m,t},\, \mathcal{V}^{\ell_2}_{\mathcal{J}^m,t})&=& Cov(\mathcal{V}^{\ell_1}_{\mathcal{J}^m+1,t},\, \mathcal{V}^{\ell_2}_{\mathcal{J}^m+1,t})+\gamma_{\mathcal{J}^m+1}^{\ell_1,\ell_2} \\
\vdots\hspace{10mm}&=&\hspace{10mm}\vdots \\
Cov(\mathcal{V}^{\ell_1}_{\mathcal{J}^m+n-1,t},\, \mathcal{V}^{\ell_2}_{\mathcal{J}^m+n-1,t})&=& Cov(\mathcal{V}^{\ell_1}_{\mathcal{J}^m+n,t},\, \mathcal{V}^{\ell_2}_{\mathcal{J}^m+n,t})+\gamma_{\mathcal{J}^m+n}^{\ell_1,\ell_2}.
\end{eqnarray}
By summation, we obtain
\begin{eqnarray}
Cov(\mathcal{V}^{\ell_1}_{\mathcal{J}^m-1,t},\, \mathcal{V}^{\ell_2}_{\mathcal{J}^m-1,t})&=& Cov(\mathcal{V}^{\ell_1}_{\mathcal{J}^m+n,t},\, \mathcal{V}^{\ell_2}_{\mathcal{J}^m+n,t}) + \sum_{j=0}^n\gamma_{\mathcal{J}^m+j}^{\ell_1,\ell_2}.
\end{eqnarray}
For the part consisting of the wavelet coefficient covariance, we have
\begin{equation}
\sum_{j=0}^n\gamma_{\mathcal{J}^m+j}^{\ell_1,\ell_2} = Cov(\mathcal{V}^{\ell_1}_{\mathcal{J}^m-1,t},\, \mathcal{V}^{\ell_2}_{\mathcal{J}^m-1,t}) - Cov(\mathcal{V}^{\ell_1}_{\mathcal{J}^m+n,t},\, \mathcal{V}^{\ell_2}_{\mathcal{J}^m+n,t}).
\end{equation}
Let us denote $s_r$ as a sum of the wavelet coefficients covariances up to scale $r$, i.e.,
\begin{equation}
s_r=\sum_{j=0}^r \gamma_j^{\ell_1,\ell_2}.
\end{equation}
Then, for any positive integer $r$ such that $r>\mathcal{J}^m$, we have:
\begin{eqnarray}
s_r &=& \sum_{j=0}^{\mathcal{J}^m-1}  \gamma_j^{\ell_1,\ell_2} + \sum_{j=0}^{r-\mathcal{J}^m} \gamma_{\mathcal{J}^m+j}^{\ell_1,\ell_2}\\
&=& Cov\left(\mathcal{V}^{\ell_1}_{\mathcal{J}^m-1,t},\, \mathcal{V}^{\ell_2}_{\mathcal{J}^m-1,t}\right) - Cov\left(\mathcal{V}^{\ell_1}_{r,t},\, \mathcal{V}^{\ell_2}_{r,t}\right) + \sum_{j=0}^{\mathcal{J}^m-1}  \gamma_j^{\ell_1,\ell_2}.
\end{eqnarray}
Hence, for any two positive integers $r_1, r_2 >\mathcal{J}^m$, we can write
\begin{equation}
\vert s_{r_1}-s_{r_2}\vert = \vert Cov\left(\mathcal{V}^{\ell_1}_{r_1,t},\, \mathcal{V}^{\ell_2}_{r_1,t}\right)  - Cov\left(\mathcal{V}^{\ell_1}_{r_2,t},\, \mathcal{V}^{\ell_2}_{r_2,t}\right) \vert.
\end{equation}
Based on the result of \cite{WhitcherGP2000} (lemma 1, page 2), for any $\epsilon>0$, there exists $\mathcal{J}^m_\epsilon$ such that for a positive integer, $r>\mathcal{J}^m_\epsilon$ holds:
\begin{equation}
\label{WhLemma1}
\vert Cov\left(\mathcal{V}^{\ell_1}_{r,t},\, \mathcal{V}^{\ell_2}_{r,t}\right)\vert < \epsilon.
\end{equation}
Then (\ref{WhLemma1}), for any $\epsilon>0$, there exists $\mathcal{J}^m_\epsilon$ such that for positive integers $r_1, r_2 > \mathcal{J}^m_\epsilon$, we obtain
\begin{equation}
\vert s_{r_1}-s_{r_2}\vert \le 2\epsilon ,
\end{equation}
As a result, the sequence $\{ s_r\}$ is Cauchy and has a limit:
\begin{equation}
\lim_{r\to\infty} s_r = \sum_{j=0}^{\infty}   \gamma_j^{\ell_1,\ell_2} = Cov\left(\mathcal{V}^{\ell_1}_{\mathcal{J}^m-1,t},\, \mathcal{V}^{\ell_2}_{\mathcal{J}^m-1,t}\right) + \sum_{j=0}^{\mathcal{J}^m-1}  \gamma_j^{\ell_1,\ell_2}.
\end{equation}
Then, it follows that
\begin{equation}
\sum_{j=\mathcal{J}^m}^{\infty}  \gamma_j^{\ell_1,\ell_2} = Cov\left(\mathcal{V}^{\ell_1}_{\mathcal{J}^m-1,t},\, \mathcal{V}^{\ell_2}_{\mathcal{J}^m-1,t}\right),
\end{equation}
which implies (c.f. \ref{Wconinf})
\begin{equation}
Cov(Y_{t_,\ell_1},Y_{t_,\ell_2}) = \sum_{j=0}^{\infty}  \gamma_j^{\ell_1,\ell_2},
\end{equation}
This completes the proof.
\qed

\subsection{Wavelet realized covariance estimator}
\label{wrcapp}
Based on quadratic covariation decomposition and wavelet covariance, let us define the wavelet realized covariance estimator of processes $(Y_{t,\ell_1},Y_{t,\ell_2})$ in $L^2(\mathbb{R}$) over a fixed time horizon $[0\le t\le T]$ as
\begin{equation}
\label{wrc}
\widehat{QV}_{\ell_1,\ell_2}^{(WRC)}=\sum_{j=1}^{\mathcal{J}^m+1} \sum_{k=1}^{N} \mathcal{W}_{j,k}^{\ell_1}  \mathcal{W}_{j,k}^{\ell_2},
\end{equation}
where $N$ is the number of intraday observations, and $\mathcal{W}_{j,k}^{\ell}$ are the intraday MODWT coefficients of the process $\Delta Y_{t,\ell} = (\Delta_1 Y_{t,\ell},\ldots,\Delta_N Y_{t,\ell})$ on scale $j$ and are unaffected by the boundary conditions. $\mathcal{J}^m\le \log_2 N$ denotes the number of scales considered. Hence, we use a $N\times \mathcal{J}^m+1$ matrix of wavelet coefficients where the first $\mathcal{J}^m$ subvectors are the MODWT wavelet coefficients at $j=1,\dots,\mathcal{J}^m$ levels, and the last subvector consists of the MODWT scaling coefficients at the $\mathcal{J}^m$ level. 

Using the results of \cite{serroukhwalden2000a, serroukhwalden2000b}, we can write $\widehat{QV}_{\ell_1,\ell_2}^{(RC)}=\widehat{QV}_{\ell_1,\ell_2}^{(WRC)}$ because the realized covariance of the zero mean return process over $[0\le t\le T]$ can be written as
\begin{equation}
\sum_{i=1}^N \Delta_i Y_{t,\ell_1} \Delta_i Y_{t,\ell_2} = \sum_{j=1}^{\mathcal{J}^m+1} \sum_{k=1}^{N} \mathcal{W}_{j,k}^{\ell_1}  \mathcal{W}_{j,k}^{\ell_2}.
\end{equation}  The estimator in Eq. (\ref{wrc}) takes the asymptotic properties of
the $\widehat{QV}_{\ell_1,\ell_2}^{(RC)}$, and the estimator converges in probability to the quadratic covariation
\begin{equation}
\widehat{QV}_{\ell_1,\ell_2}^{(WRC)}\overset{p}{\rightarrow}{QV_{\ell_1,\ell_2}}.
\end{equation}

\section{Bootstrapping the co-jumps}
\label{sec:bootstrap}

Under the null hypothesis of no jumps and co-jumps in the $(Y_{t,\ell_1},Y_{t,\ell_2})$ process,
\begin{eqnarray}
\mathcal{H}^0:\widehat{QV}^{(RC)}_{\ell_1,\ell_2}-\widehat{IC}^{(JWC)}_{\ell_1,\ell_2} &=& 0  \\
\mathcal{H}^A:\widehat{QV}^{(RC)}_{\ell_1,\ell_2}-\widehat{IC}^{(JWC)}_{\ell_1,\ell_2} &\ne& 0.
\end{eqnarray}
We propose a simple test statistic that can be used to detect significant co-jump variation. If a significant difference exists between the quadratic covariation and integrated covariance, then it is highly probable that we will observe a co-jump variation, possibly because of co-jump(s) or large disjoint jump(s). In this case, the $\mathcal{H}^0$ is rejected for its alternative.

When the null hypotheses of no jumps holds,
$\widehat{QV}^{(RC)}_{\ell_1,\ell_2}-\widehat{IC}^{(JWC)}_{\ell_1,\ell_2}$
is asymptotically independent from
$\widehat{QV}^{(RC)}_{\ell_1,\ell_2}$ conditional on the volatility path, and we can use two independent random variables to set the Hausman-type statistics to test for the presence of jumps. We proceed by scaling $\widehat{QV}^{(RC)}_{\ell_1,\ell_2}-\widehat{IC}^{(JWC)}_{\ell_1,\ell_2}$ by the difference in the variances of both estimators, which we obtain using a bootstrap procedure.

Under the null hypothesis of no jumps and co-jumps, we generate $i$ intraday returns $(r_{i,\ell_1}^*,r_{i,\ell_2}^*)$ with integrated covariance determined based on empirical estimates as
\begin{eqnarray}
r_{i,\ell_1}^*&=&\sqrt{\frac{1}{N} \widehat{IC}^{(JWC)}_{\ell_1,\ell_1}}\eta_{i,\ell_1} \\
r_{i,\ell_2}^*&=&\sqrt{\frac{1}{N}\widehat{IC}^{(JWC)}_{\ell_2,\ell_2}}\left( \widehat{\rho}_{\ell_1,\ell_2} \eta_{i,\ell_1}+\sqrt{1- \widehat{\rho}^2_{\ell_1,\ell_2}} \eta_{i,\ell_2}\right), \\
\end{eqnarray}
with $ \widehat{\rho}_{\ell_1,\ell_2}$ being the correlation obtained
from the $\widehat{\bm{IC}}^{(JWC)}$ matrix, and $\eta_{i,\ell_1} \sim \mathcal{N}(0,1)$ and $\eta_{i,\ell_2} \sim \mathcal{N}(0,1)$.
Now, we use $(r_{i,\ell_1}^*,r_{i,\ell_2}^*)$ to compute $\widehat{QV}^{(RC)*}_{\ell_1,\ell_2}$ and $\widehat{IC}^{(JWC)*}_{\ell_1,\ell_2}$. Generating $b=1,\ldots,B$ realizations, we obtain $\mathcal{Z}^*=(\mathcal{Z}^{(1)},\mathcal{Z}^{(2)},\ldots,\mathcal{Z}^{(B)})$ as
\begin{equation}
\mathcal{Z}^*=\frac{\widehat{QV}^{(RC)*}_{\ell_1,\ell_2} - \widehat{IC}^{(JWC)*}_{\ell_1,\ell_2}}{\widehat{QV}^{(RC)*}_{\ell_1,\ell_2}}.
\end{equation}
which can be used to construct a bootstrap statistic to test the null hypothesis of no co-jumps as
\begin{equation}
\mathcal{Z}=\frac{\frac{\widehat{QV}^{(RC)}_{\ell_1,\ell_2} - \widehat{IC}^{(JWC)}_{\ell_1,\ell_2}}{\widehat{QV}^{(RC)}_{\ell_1,\ell_2}}-E(\mathcal{Z}^*)}{\sqrt{Var(\mathcal{Z}^*)}}\sim\mathcal{N}(0,1).
\end{equation}
The bootstrap expectation and variance depend on the data. We rely on
the assumptions of \cite{dovonon2014bootstrapping}. Thus, by identifying days when the co-jump component is present, we can estimate the off-diagonal elements of the covariance matrix $\widehat{\bm{IC}}^{(JWC*)}$ as
\begin{equation}
\label{iccxy}
\widehat{IC}_{\ell_1,\ell_2}^{(JWC*)} = \widehat{QV}^{(RC)}_{\ell_1,\ell_2}\mathbbm{1}_{\{ \vert\mathcal{Z}\vert \le \phi_{1-\alpha/2}\}}+\widehat{IC}^{(JWC)}_{\ell_1,\ell_2}\mathbbm{1}_{\{ \vert\mathcal{Z}\vert > \phi_{1-\alpha/2}\}},
\end{equation}
where $\phi_{1-\alpha/2}$ is a critical value for the two-sided test
with a significance level $\alpha$. Finally, we estimate all elements of the (continuous) covariance matrix:
\begin{equation}
\widehat{\bm{IC}}^{(JWC*)}=\left(
\begin{array}{cc}
\widehat{IC}^{(JWC*)}_{\ell_1,\ell_1} & \widehat{IC}_{\ell_1,\ell_2}^{(JWC*)}\\
\widehat{IC}_{\ell_2,\ell_1}^{(JWC*)} & \widehat{IC}^{(JWC*)}_{\ell_2,\ell_2}
\end{array}
\right).
\end{equation}

\end{document}